# Recent advances in blood rheology: A review

Antony N. Beris,*[a] Jeffrey S. Horner,[ab] Soham Jariwala,[a] Mathew J. Armstrong[c] and Norman J. Wagner[a]

Due to the potential impact on the diagnosis and treatment of various cardiovascular diseases, work on the rheology of blood has significantly expanded in the last decade, both experimentally and theoretically. Experimentally, blood has been confirmed to demonstrate a variety of non-Newtonian rheological characteristics, including pseudoplasticity, viscoelasticity, and thixotropy. New rheological experiments and the development of more controlled experimental protocols on more extensive, broadly physiologically characterized, human blood samples demonstrate the sensitivity of aspects of hemorheology to several physiological factors. For example, at high shear rates the red blood cells elastically deform, imparting viscoelasticity, while and at low shear rates, they form "rouleaux" structures that impart additional, thixotropic behavior. In addition to the advances in experimental methods and validated data sets, significant advances have also been made in both microscopic simulations and macroscopic, continuum, modeling, as well as novel, multiscale approaches. We outline and evaluate the most promising of these recent developments. Although we primarily focus on human blood rheology, we also discuss recent observations on variations observed across some animal species that provide some indication on evolutionary effects.

## Introduction

Blood is an especially important biofluid, critical for life, through a multiplicity of roles in the cardiovascular circulation, from transferring oxygen and nutrients to the tissues, to removing carbon dioxide and waste products of metabolism. Evidence of the importance of cardiovascular circulation comes from the fact that cardiovascular diseases continue to present the leading cause of death worldwide.[1] Historically, blood has played a very important role in medicine, starting at least with Hippocrates—see[2] for a historical perspective. As a result of millions of years of evolution in perfecting its role in cardiovascular circulation, blood has evolved to become an especially complex fluid that is species dependent. Broadly speaking, it is comprised of a concentrated suspension of large, deformable, red blood cells (RBC, erythrocytes), (~ 6-8 μm diameter with a biconcave shape), additional, larger, spherical white cells (leukocytes ~ 7-22 μm), and many, smaller (~ 2-4 μm), oval-shaped platelets within plasma which is an aqueous solution of a variety of proteins.[3, 4] As the volume fraction of RBCs (hematocrit) by far exceeds that of any other component (see also **Table 1**), it is common to consider blood as a suspension of flexible RBCs in an otherwise Newtonian fluid, as will be discussed. Complex interactions between RBCs result in aggregates (rouleaux) that can grow to form a temporary network at low shear rates,[5] which results in a significantly complex rheological behavior.[2] Consequently, hemorheology is characterized by a shear thinning viscosity,[6] a general viscoelastic response to transient flow deformations[7] and a non-zero yield stress[8] with associated thixotropy[9] – see, for example, a recent review[2] and references therein.

The objective of the present review is, by presenting additional, more recent, references and an enhanced emphasis on the microscopic characteristics of blood, to provide an updated description of blood rheology. We show how as a result of a better understanding of blood's microstructure, and of the development of more appropriate experimental protocols and rheological experiments, more accurate models with fewer parameters and better physical meaning can be developed. Such models not only lead to better flow predictions, but also provide a strong foundation on which the modeling of the rheology of other biofluids can be built, thereby significantly extending our understanding of soft matter encountered in living organisms in general.

In the following, we provide additional information on the underlying blood microstructure and how is this affected by the flow, on the experimental methodologies and on new theoretical developments at various levels of description. Models derived from these advances range from simpler, generalized Newtonian, approaches, suitable mostly for steady-state shear-dominated flows, to thixotropic, elastoviscoplastic models, appropriate to study blood's complex transient rheological behavior. The emphasis of this review is continuum models to describe bulk blood rheology, although important connections to microscopic and multiscale models are also mentioned. Importantly, developments at the microscopic level, addressing nonhomogeneous (stress-induced migration and syneresis) effects are also presented, along with emerging

[a] *Center for Research in Soft Matter and Polymers, Department of Chemical and Biomolecular Engineering, University of Delaware, Newark, DE 19716.*
[b] *Current address: Thermal/Fluid Component Sciences Department, Sandia National Laboratories, Albuquerque, NM 87123*
[c] *Department of Chemistry and Life Science, Chemical Engineering Program, United States Military Academy, West Point, NY 10996*

**E-mail: beris@udel.edu*

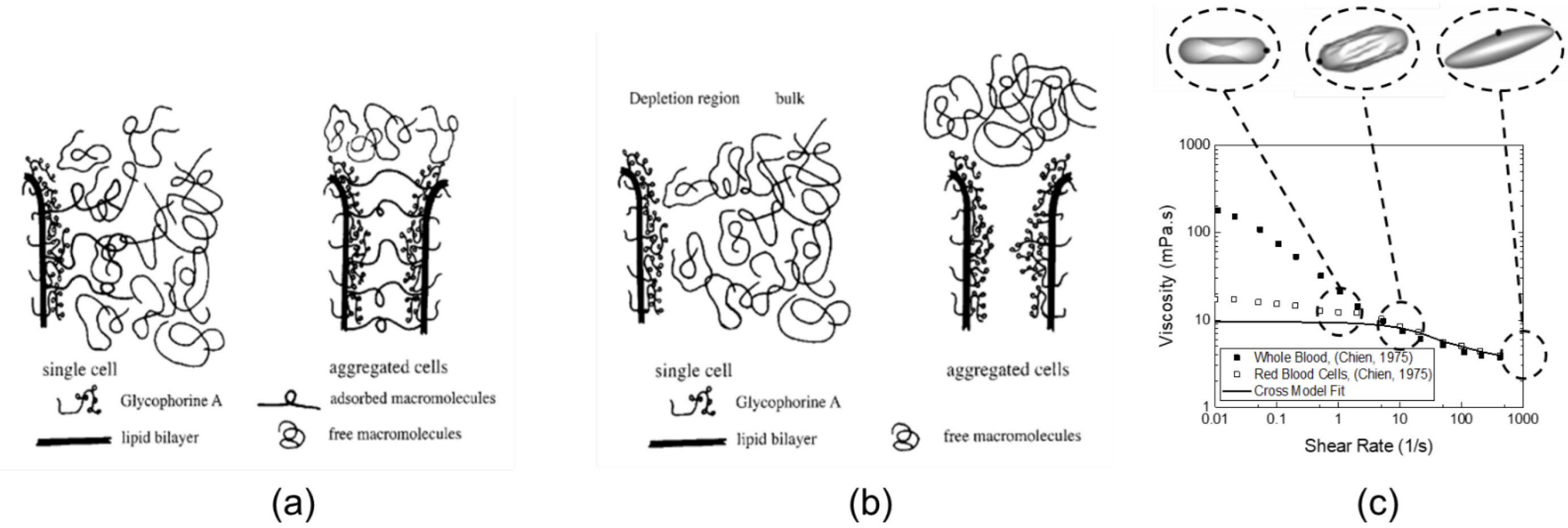

**Fig. 1** A qualitative depiction of the two hypothesized mechanisms for RBC aggregation. The bridging model (a) suggests that macromolecules adsorb to the surface (lipid bilayer) of RBCs and link to adjacent RBCs. The depletion model (b) suggests that depletion layers form near RBC membranes, and when two depletion layers overlap, osmotic pressure drives the RBCs together. Figure taken from Baumler et al., 1999.[19] (c) Viscosity of whole blood (solid symbols) compared to that of a suspension of purified red blood cells (open symbols) fit with the Cross model.[52] Weak attractive forces drive rouleaux formation and yield stress development. The images of typical RBC shape show the shear-induced shape deformation driving shear thinning at high shear rates. Bottom figure obtained from Chien, The Red Blood Cell, Vol. 2, Academic Press, 1975[5] and deformed RBC depictions obtained from Omori et al., Physical Review E, 2012,[10] reprinted with permission.

new hybrid micro-macro approaches. Also of interest, new comparative blood rheology results from various animal species are discussed. Those provide an indication of the effect of evolution, of importance as we discuss biofluids presenting a clear distinction from other, man-made, soft matter fluid behavior. Finally, we conclude with an overall assessment as to where the current state of understanding of blood rheology stands along with some thoughts on potential new developments in the near future.

## Hemorheology characteristics

**Table 1** Characteristic scales in blood rheology. Data from Supplemental material of Horner et al., Journal of Rheology, 2019[13] and Horner, 2020.[14] Additional details and physical basis of the parameters are presented in the "constitutive blood modeling" section.

| Parameter | Description | units | Average value | Range |
|---|---|---|---|---|
| $H$ | hematocrit | % | 42 | [38.5 - 46] |
| $c_f$ | fibrinogen | g/dL | 0.252 | [0.210 - 0.290] |
| $\eta_0$ | Cross model zero-shear viscosity | Pa s | 0.0082 | [0.005 - 0.010] |
| $\eta_\infty$ | Cross model infinite-shear viscosity | Pa s | 0.0035 | [0.0029 - 0.0041] |
| $\eta_p$ | plasma viscosity | Pa s | 0.0013 | [0.001 - 0.0016] |
| $\sigma_y$ | yield stress | Pa | 0.0024 | [0.0018 - 0.0040] |
| $\eta_R$ | rouleaux viscosity | Pa s | 0.041 | [0.020 - 0.060] |
| $\tau_c = \eta_c/G_c$ | Free cell relaxation time | s | 0.0139 | [0 - 0.04] |
| $\tau_R = \eta_R/G_R$ | Rouleaux relaxation time | s | 0.1138 | [0 - 0.3] |
| $\tau_\lambda$ | Brownian relaxation time | s | 1.4645 | [1.1 – 1.8] |

The most obvious non-Newtonian characteristic of blood rheology is its shear thinning in steady shear flow (see **Fig. 1**(c))[5, 10] and has been studied extensively for more than 100 years.[11, 12] However, with time, people discovered that blood rheology exhibits additional more complex rheological characteristics, including viscoplasticity, viscoelasticity and thixotropy. To aid in this discussion, in **Table 1** we provide a list of characteristic values (reproduced from Horner et al.[13] and Horner[14]) of a selection of various quantities that can be used to characterize various aspects of hemorheology, such as various measures for limiting values (at zero and high shear rates) of blood viscosity, and its contribution from plasma and rouleaux, characterizing shear thinning, yield stress characterizing viscoplasticity, and three typical characteristic times useful to characterize two separate components of viscoelasticity (originating from the free RBC deformation, rouleaux aggregates) and thixotropy associated with rouleaux formation due to Brownian motion. Using these characteristic time values and shear rates of 1, 0.01, 1000 $s^{-1}$, we provide in **Table 2** typical, low and high range values, respectively, for a selected set of important dimensionless groups. The cell capillary number $Ca_G$ is based on a definition involving the shear modulus of the red blood cells offered by Kumar and Graham[15] and Liu et al.[16] Others have used a similar definition based on the RBC membrane Young's modulus.[17] The tabulated values correspond to a characteristic frequency of $1\ \mathrm{Hz}$ ($\omega \equiv 2\pi f = 6.28\ \mathrm{rad/s}$) and involve vessel radii values corresponding to capillaries (low value), intermediate value corresponding to Womersley value of 1 (typical value), and aorta (high value). The Bingham number is based on wall shear stress values as far as typical and higher values are concerned (those change little as we move from an average arteriole, corresponding to Womersley number of 1 (typical value) to the capillaries (high value)). However, a low range value of 0 is given considering that the stress decreases and reaches zero as the centerline is approached. Moreover, with this opportunity we need to mention that due to blood's thixotropic characteristics, the Bingham number based on blood's zero shear rate yield stress (what is used here) is of little physical meaning when its reported value is as high as shown here as the effective yield stress quickly disappears with the breakup of the rouleaux structures at the higher stresses. In contrast, it is meaningful and of importance when rheological measurements at low shear rates are performed with a low Bingham number (below or at order one) corresponding to conditions under which sedimentation may become important. The values for Deborah number and Weissenberg number are based on the characteristic viscoelastic timescales. The high range for Deborah number is based upon the higher harmonics of the pulsatile flow.

**Table 2** Characteristic dimensionless groups and their typical values

| Group | Expression | Typical Values | Range |
|---|---|---|---|
| Péclet number | $\mathrm{Pe} = \frac{6\pi\eta a^3}{k_B T}\dot{\gamma}$ | $3.5 \times 10^3$ | $[50 - 3.5 \times 10^5]$ |
| RBC capillary number | $\mathrm{Ca_G} = \frac{\eta\dot{\gamma}a}{G_m}$ | $1.4 \times 10^{-2}$ | $[2 \times 10^{-4} - 1.4]$ |
| Womersley number | $\mathrm{Wo} = r\left(\frac{\omega\rho}{\eta}\right)^{1/2}$ | 1 | $[0 - 13]$ |
| Bingham number | $\mathrm{Bi} = \frac{\sigma}{\sigma_y}$ | 900 | $[0 - 1500]$ |
| Deborah number | $\mathrm{De} = \tau\omega$ | 0.1 | $[0 - 20]$ |
| Weissenberg number | $\mathrm{Wi} = \tau\dot{\gamma}$ | 0.01 | $[1 \times 10^{-3} - 10]$ |

For comparison, in **Fig. 1**(c)[5, 10] taken from the work by Chien[5] are measurements on hardened RBCs and RBCs suspended in an albumin-Ringer solution in which no RBC aggregation should occur. As shown by depiction of RBC shape, the strong shear thinning response arises primarily because of the formation and breakup of the aggregated RBC rouleaux structures. Comparatively, the rheological data on RBCs in the absence of proteins demonstrates a less pronounced shear thinning behavior. The shear thinning seen for this curve arises from the deformable nature of the RBCs. At low shear rates, RBCs will rotate in a motion similar to tank treading which results in a decreased viscosity from the hardened case. The viscosity will then decrease further at high shear rates as the RBCs begin to deform.[18]

While it is known that plasma proteins, particularly fibrinogen, are required to achieve rouleaux formation within the blood, the exact mechanism driving the aggregation of RBCs is still debated. At present there are two prevailing theories for the forces driving the reversible RBC aggregation. The first hypothesis is "bridging" by adsorbed proteins, as pictured in **Fig. 1**(a).[19, 20] Macromolecules adsorb onto the surfaces of adjacent RBCs and thereby physically crosslink them into coin-stack structures known as rouleaux.[21, 22] This hypothesis requires that the adsorption of the fibrinogen onto the RBC be sufficiently strong to overcome the loss of entropy as well as the electrostatic repulsive forces between the RBCs in close proximity, yet weak enough to enable reversible separation at increased shear stresses. The competing hypothesis, pictured in **Fig. 1**(b),[19, 20] suggests the opposite effect. Instead of macromolecules being adsorbed onto the RBC membranes, a "depletion" layer exists which arises from steric hindrance and loss of configurational entropy for macromolecules near the surface of the RBC. The overlap of two depletion layers will result in an osmotic pressure gradient which can drive the two RBC surfaces together.[23] This hypothesis was formally proposed before the bridging hypothesis, but experimental support for this hypothesis did not come until much later.[19]

The bridging hypothesis is supported by numerous works. Notably, *in vitro* experiments were performed on RBC aggregation in dextran solutions of differing molecular masses. It was found that with increasing molecular mass, the RBC separation distance between adjacent RBCs within the rouleaux also increased. Additionally, with increased molecular mass, the concentration of dextran required to facilitate aggregation was decreased.[24] This result supports the bridging hypothesis because with increased separation distance, caused by increased size of the macromolecules, fewer macromolecules would need to adsorb to overcome increasingly weaker electrostatic repulsive forces. More recently, experiments performed on an isolated RBC aggregate moved between solutions of differing protein concentrations also support the bridging hypothesis, as it was shown that the aggregate depends on both the initial solution and the final solution. This contrasts with the depletion hypothesis that would suggest that only the final solution should determine the aggregation tendency.[25]

There is also experimental evidence supporting the depletion hypothesis. Particularly, it was found that it is the hydrodynamic radius of macromolecules, instead of the polymer type, that has a strong relation to the aggregation tendencies of the RBCs.[26] If the bridging hypothesis were true, then the specific type of macromolecules, in addition to the hydrodynamic radii, would have affected the aggregation tendencies. Additionally, unlike the bridging hypothesis, the depletion hypothesis does not require a dynamic readjustment of adsorbed macromolecules to facilitate the reorientation of the RBCs into the rouleaux.

In addition to a pronounced shear thinning behavior, the presence of rouleaux in blood also gives rise to viscoelasticity and thixotropy. Due to the weak viscoelasticity of blood and a transition to nonlinear behavior at low strain amplitudes, the linear viscoelastic behavior of blood is typically difficult to measure. Thus, rheological techniques, such as large amplitude oscillatory shear (LAOS), which measure the thixotropy and viscoelasticity simultaneously, are more commonly used.[27] At low shear rates, where the rouleaux are present, there is considerably more elasticity present in the sample; whereas, at high shear rates, the elasticity present is more representative of that of the isolated RBCs. The thixotropic behavior of blood arises from the structural evolution of the rouleaux which introduces a thixotropic time scale. A common technique, pioneered by Bureau and coworkers[9], for measuring this is the triangular ramp test, often termed a thixotropic loop test[28]. For this test, the shear rate increases linearly from zero over a certain time then decreases linearly over the same time back to rest. Thixotropy manifests as a stress hysteresis loop containing a crossover; whereas, viscoelastic effects will manifest only as a loop.[9]

The complex rheological properties of blood are often reported in an (overly) simplified manner by a few characteristic metrics. One such metric which governs the low shear behavior of blood is the yield stress. The yield stress of blood is typically very small, on the order of 1 mPa, and consequently difficult to measure, and can be easily missed if one is not careful to avoid the wall slip phenomenon.[29] The most common method for determining the yield stress of blood is through extrapolation of the steady shear data to zero shear rate using a model, as will be discussed in the sections to follow. An example of the consequences of wall slip in rheological measurements of blood viscosity is shown in **Fig. 2** taken from Picart et al.[29] Creep measurements, another method of determining the yield stress for a fluid, are typically not accurate enough given the low stresses required.

Adding to the difficulty of accurately measuring the yield stress is the development of a cell depletion wall near the wall that is even more evident when measuring blood at low shear rates. This phenomenon arises due to a syneresis effects where RBCs will migrate towards the center of the measurement device leaving a plasma enriched in platelets layer near the walls. This effect can be seen for pressure driven flow in a capillary in **Fig. 3**(a)[30] and derived from magnetic resonance microscopy (MRM) velocity measurements of blood in a Couette cell, as shown in **Fig. 3**(b).[31] The formation of this layer leads to an exponentially decreasing apparent viscosity with time as the plasma region near the walls develops and has a significantly lower viscosity than the bulk region. Thus, when measuring blood at low shear rates, it is often advisable to extrapolate this effect back to zero or take the maximum stress measurement.[32] This effect is particularly pronounced at low shear rates as the aggregate sizes are larger and the difference in viscosity between the bulk and the depletion layer is more significant. At the relatively fast flow rates encountered in Poiseuille flow the wall effect is much less pronounced except for the smaller arteriole and the capillary vessels due to the smaller size of the individual RBC that are the only ones present under those conditions. This effect is referred to as the Fåhræus effect or the Fåhræus-Lindqvist effect, the former referring to the decrease in local hematocrit and the latter referring to the decrease in the apparent viscosity.[33, 34]

While most studies are typically performed on whole blood, several authors have also measured the rheology of the plasma. Most studies agree that plasma is weakly viscoelastic in nature with some reports suggesting that the viscoelasticity of plasma may be especially significant in extensional flow.[35-37] However, extensional flow measurements are significantly more difficult to perform on such a low viscosity fluid and when using a filament stretch rheometer, often subject to complications arising from interfacial effects as proteins dynamically absorb to the air water interface. Therefore, there is concern with the interpretation of the observed results in extensional flow for blood[36] and how important extensional viscoelasticity is for plasma.[37]

The main physiological determinants of whole blood rheology tend to be the hematocrit and the fibrinogen concentration, with the latter affecting the aggregation tendency of the rouleaux. Various authors suggest that the low shear behavior of blood is dependent on both these parameters, with aggregation occurring only above a certain critical hematocrit.[21, 38] More recently, attention has been given to additional physiological parameters. Specifically, it has been suggested that the ratio of cholesterol levels may be important to the rheology of blood, with increased high-density lipoproteins (HDL) correlating to a decrease in the yield stress while increased low-density lipoproteins (LDL) and triglycerides lead to the inverse phenomenon.[39] It is intriguing to note here that HDL is commonly known as "good" cholesterol while LDL is commonly termed as "bad" with regard to deposition on arterial walls. There is actually some evidence from medical recommendations that the overall cholesterol effect to cardiovascular health is dependent primarily from the HDL/LDL or HDL/TL (TL=total cholesterol) ratio[40], which is therefore consistent to the findings of this preliminary rheological study[39] the results of which are further reinforced from the hypothesis that cholesterol and triglycerides (that have also been studied there) affect health primarily because they modify blood's viscosity.[41] However, caution needs to be exercised due to the small size and possibility of surface effects contamination of the experimental data on which the above preliminary rheological study was based upon.[42] Several other properties of the blood have also been suggested to affect the bulk rheology including the anticoagulant,[43] gas concentrations,[44] and blood type.[45] However, the work in these areas is sparser and will not be discussed here.

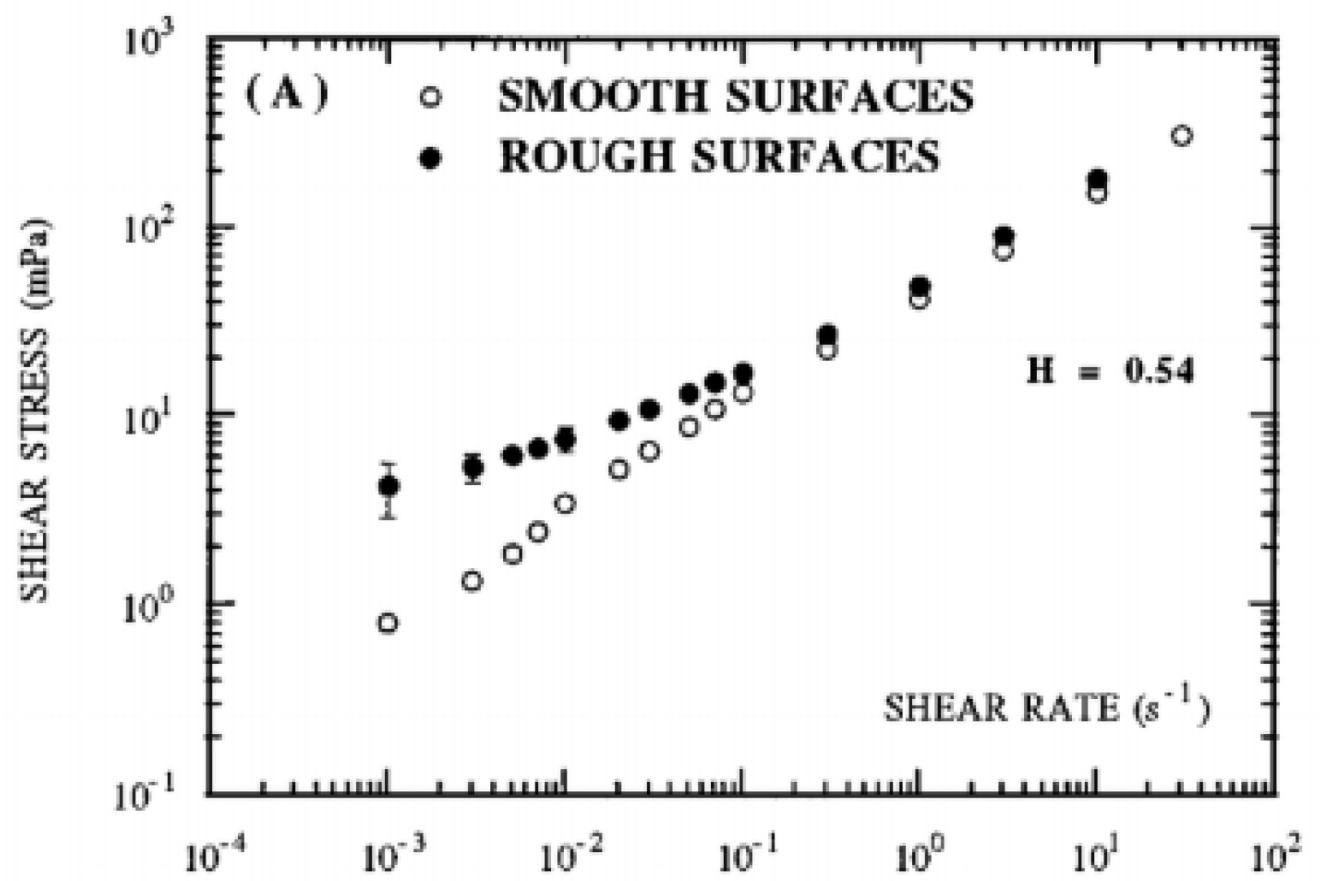

**Fig. 2** Steady shear stress measurements at varying shear rates on human blood with a hematocrit of 54%. A comparison of smooth and rough surfaces is shown, and the yielding behavior of blood can be clearly seen for the rough surfaces. Figure taken from Picart et al., Journal of Rheology, 1998,[29] reprinted with permission.

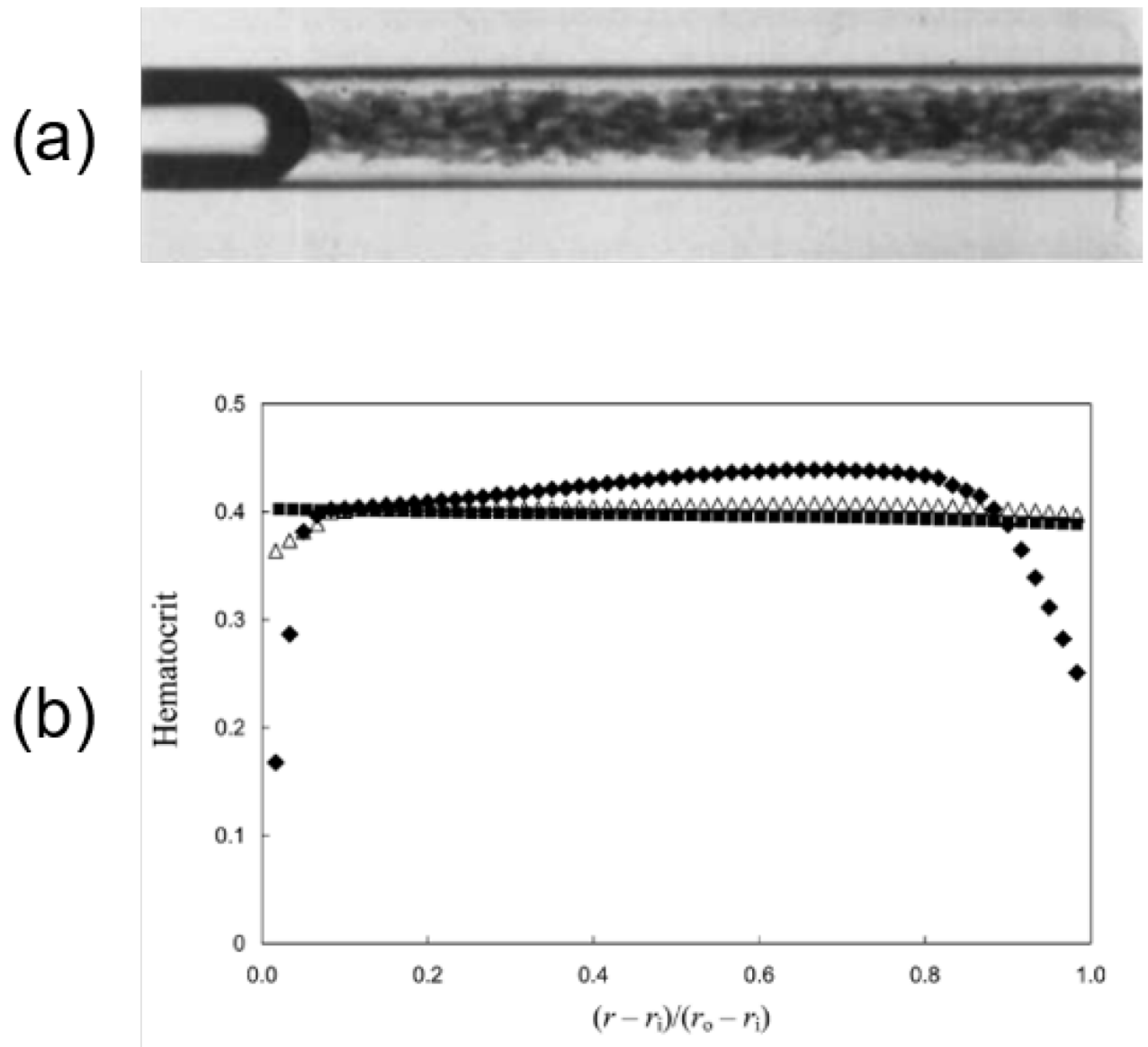

**Fig. 3** Human blood flowing in a capillary (a), where the depletion layer near the walls of the capillary can be seen. Human blood hematocrit profile across the radial gap for flow in a Couette device (b), deduced from MRM velocity measurements. Note the decreased hematocrit near the edges of the device. Figure taken from Fåhræus, Acta Medica Scandinavica, 1958[30] (a) and from Cokelet et al., Biorheology, 2005[31] (b), reprinted with permission.

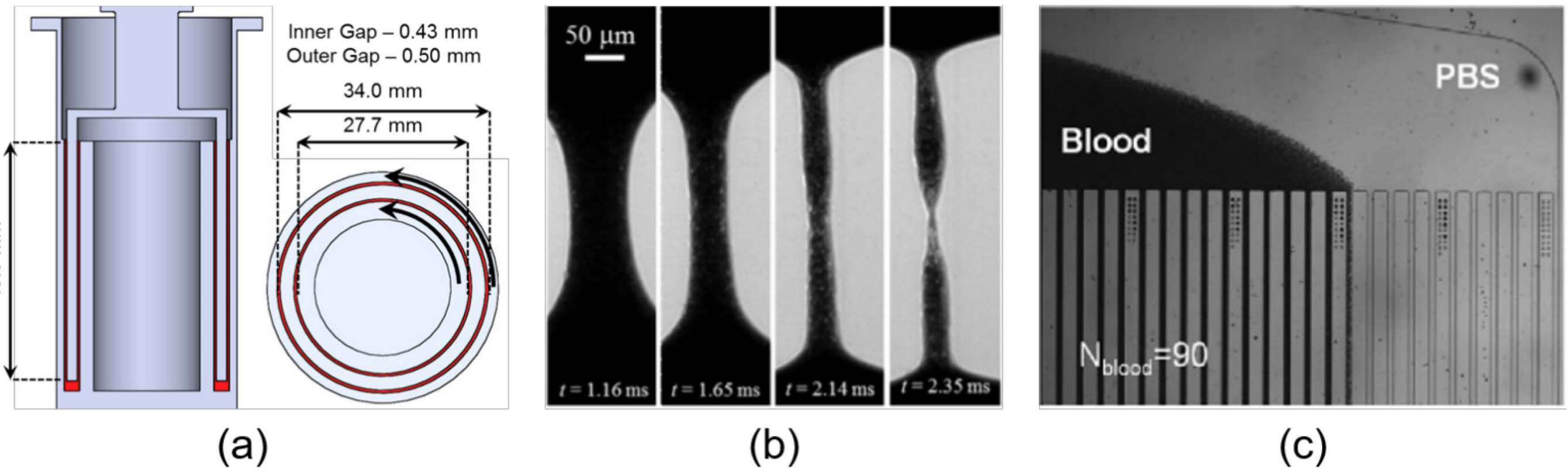

**Fig. 4** A selection of methods for measuring the bulk material properties of blood. (a) A schematic of the double wall concentric cylinder geometry used for a bulk rheometer. (b) A CaBER uniaxial extension experiment for blood immersed in oil. (c) A microfluidic viscometer which measures the viscosity of blood relative to a phosphate buffered saline (PBS) solution by evaluating the fraction of parallel capillaries that each fluid flows through for given inlet flow rates. Figures taken from Horner et al., Journal of Rheology, 2018[52] (a); Sousa et al., Journal of Rheology, 2018[36] (b); and Kang et al., Artificial Organs, 2010[59] (c), reprinted with permission.

## Experimental methods

Rheometry was historically and now, is commonly used to measure the material properties of blood. Already in the early 20th century, the industrial rheologist Scott Blair[46] described flow curves obtained from capillary data for humans and animal data with the Casson model.[2, 47]

$$\sigma_{12}^{1/2} = \sigma_{y,c}^{1/2} + (\eta_\infty \dot{\gamma})^{1/2}, \quad (1)$$

where $\sigma_{12}$ is the shear component of the stress, $\sigma_{y,c}$ is the Casson model yield stress, $\dot{\gamma}$ is the shear rate and $\eta_\infty$ is the corresponding viscosity.[38] Importantly, these early investigations into steady shear flow already identified the complex nature of blood rheology manifesting as yielding and shear thinning and suggested that a more complete rheological characterization of blood's dynamic rheology is warranted given the pulsatile nature of natural blood circulation in the arterial system—see also a recent review by Sousa et al.[48]

Accurate and reproducible measurements of blood rheology are complicated by the "living" nature of blood, its interactions with interfaces and propensity to coagulate, and the aforementioned wall cell-free layer, especially at low shear rates. Currently, the double-wall concentric cylinder, shown in **Fig. 4**(a), is the preferred geometry for standard shear rheometry as it maximizes the measurement area and vertical length while minimizing free surface area and sample volume.[49] Titanium and stainless steel cells enable proper cleaning and have minimal interaction with plasma proteins. The use of a solvent trap, temperature control, and suitable preconditioning protocol (typically shearing at a moderate to high shear rate for 30-60 s) to remove memory effects is necessary. The addition of an anticoagulant during blood withdrawal, preferably EDTA, is also required unless the study specifically focuses on the coagulation of blood.[32, 43] A variety of tests may be performed on the blood, with steady shear being the most conventional.[8, 12, 29, 50] Additional measurements have been reported in literature for small amplitude oscillatory shear (SAOS),[51] triangular ramp,[9] LAOS,[27] step shear change,[13] and unidirectional large amplitude oscillatory shear (UD-LAOS)[52] experiments, with each probing the sample under a different transient deformation. Note that UD-LAOS was conceived to better represent the kinematics of *in vivo* pulsatile blood flow in the arteries. When performing rheological measurements on blood, it is also important to conduct the measurements in a short time from withdrawal as various biological changes, such as ATP deprivation, and physical changes, such as protein adsorption to interfaces, can occur in as little as 4 hours from withdrawal.[49] Further, cooling blood between acquisition and measurement can have a hysteretic effect that persists long after the blood is reheated to body temperature.[49]

Measuring the non-shear components of the stress tensor for blood is typically more difficult. In the past, an unsuccessful attempt was conducted to measure the normal stress under shear flow for a cone and plate device.[53] Thus, it is easier to probe the additional stress components under extensional flow. This can typically be performed using either a CaBER technique or a microfluidic hyperbolic contraction. In the former, the sample is placed between two plates and stretched uniaxially while the sample diameter is monitored with a high-speed camera. This process is shown for human blood measurements in **Fig. 4**(b).[36] These measurements are very susceptible to interfacial effects, and as a result, some authors have proposed immersing the sample in an oil solution.[36] For microfluidic flow through a hyperbolic contraction, strong extensional flows are formed naturally by the channel geometry, and the pressure drop and/or velocity pathlines are monitored. While this technique has been used successfully for blood analog solutions and diluted blood,[54, 55] conducting these experiments on whole blood is more difficult due to the Fåhræus effect and the high volume fraction of RBCs.

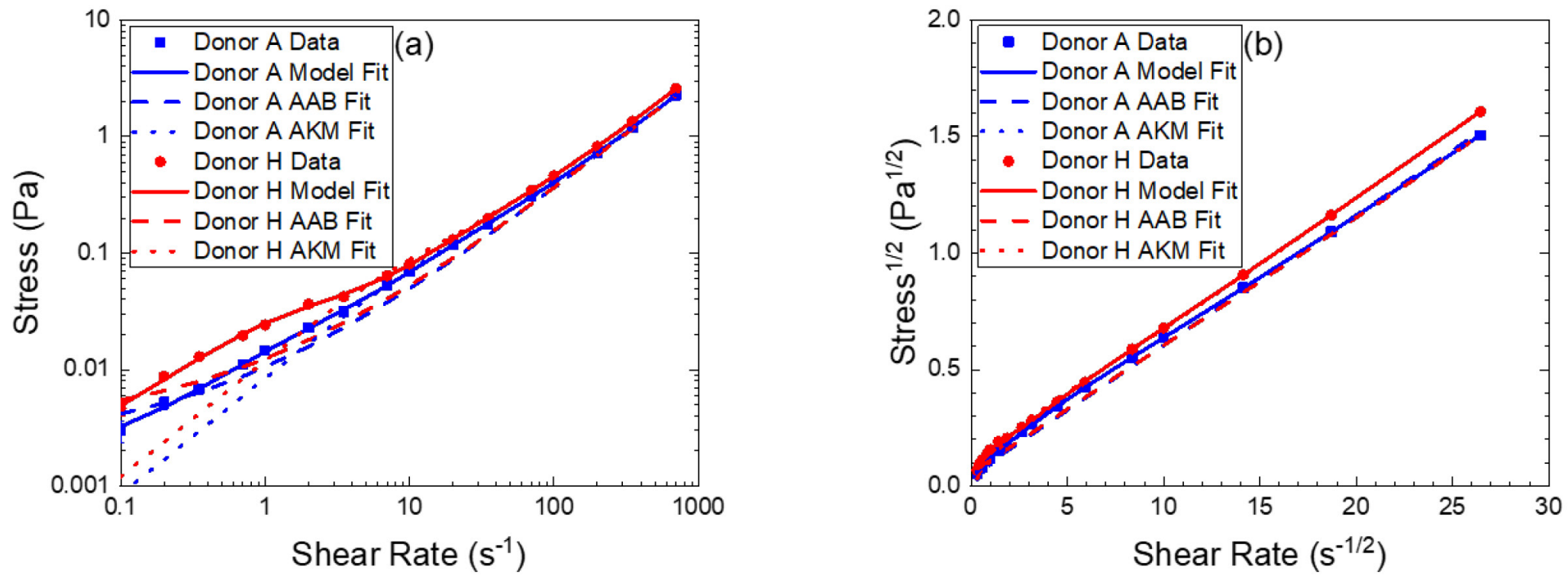

**Fig. 5** The initial steady shear data (symbols) taken on blood from Donors A and H (a) in standard coordinates and (b) Casson coordinates. Comparison of the HAWB model fit (solid lines) with the AAB model (dashed lines) and the AKM model (dotted lines) shows an improved fit which is specifically able to account for yield stress behavior and deviations from the Casson model at low shear rates. Figure adapted from Horner et al., Journal of Rheology, 2018,[52] reprinted with permission.

Other methods can also be used to measure the bulk properties of blood. Notably, viscometry has been used in the past. This is typically performed in a capillary-type viscometer where either the pressure drop or volumetric flow rate is controlled and the other is measured. These measurements are advantageous because of they are easy to perform, inexpensive and require a small amount of blood. However, it is well known that capillary viscometry of blood is subject to the Fåhræus effect whereby the apparent viscosity decreases with the diameter of the capillary. Additionally, the pressure drops will typically be small, and thus, it is difficult to measure at low flow rates. Lastly, these measurements also require *a priori* knowledge of the steady shear viscosity profile, and hence, a constitutive model is typically assumed.

Microfluidics has also been employed to measure the viscosity of blood. Various configurations have been suggested in literature including a droplet based device,[56] a gravity driven device for flow in a rectangular geometry,[57] a flow through a slit device,[58] and a multiple capillary in parallel device, the last pictured in **Fig. 4**(c).[59] While these devices can typically measure blood viscosity with a very small amount of blood, their major limitation tends to be the inability to reach low shear rates. Additionally, if any free surfaces are present, interfacial effects may affect the measurements significantly. The dynamic moduli values associated with blood's linear viscoelasticity can be determined through single particle tracking passive microrheology.[60] These experiments involve tracking the diffusion of a tracer particle over time, but are limited by the opaqueness of heterogeneous nature of blood.

For dynamic blood measurements, both *in vivo* flow conditions and *in vitro* techniques such as bulk rheology or microchannel flow can be combined with other techniques to obtain spatial velocity measurements. Two of the more common techniques used are particle image velocimetry (PIV) or particle tracking velocimetry (PTV).[58] These involve tracking particles dispersed in the blood with a high-speed camera then using a computer program to obtain velocity measurements and particle trajectories. For blood, standard tracer particles will typically aggregate, leading to poor measurements. One technique to overcome this limitation is to track the individual RBCs; however, this often requires dilution of the sample.[61] Another technique is to fluorescently label a portion of the RBCs so they can be better tracked.[62] A third technique is to coat standard tracer particles with polyethylene glycol (PEG) to create hydrophobic PEGylated particles which will not aggregate.[63] Alternative to PIV/PTV techniques, MRM[64] and laser Doppler velocimetry (LDV)[65] measurements can also both be used to track the velocity of blood. The analysis and interpretation of the microfluidic measurements can greatly benefit from the recent advances in microscopic simulations of blood.[66]

The aggregation tendency of RBCs can be more directly measured through alternative techniques. One of the more common techniques for this is microscopy. Typically, the aggregation tendency is quantified through an aggregation index. This index is a ratio of the total number of units in the aggregated blood sample including both aggregates and freely suspended individual RBCs to the total number of RBCs in the imaged sample.[24] The morphology of the aggregates can also be quantified through an aggregate shape parameter (ASP), which can be computed using various computer image analysis programs.[67] Microscopy has to be coupled with other dynamic techniques such as shear rheometry,[68] pressure driven flow in a microchannel,[69] or extensional rheometry[36] to give a better understanding of how the RBC aggregates change under flow. However, depending on the condition, the sample may need to be diluted to better visualize the individual RBC aggregates. Alternatively, if the sample is too concentrated, light

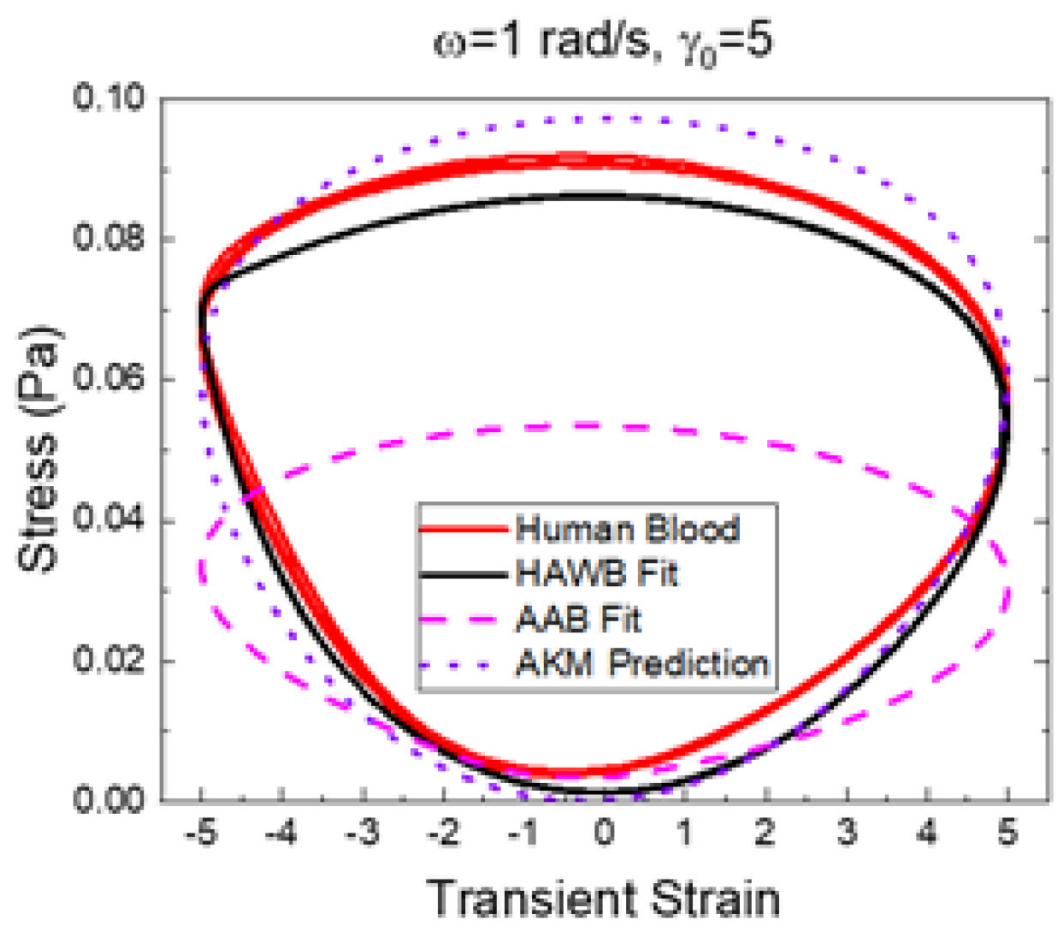

**Fig. 6** UD-LAOS experimental data for human blood at a frequency of 1 rad/s and a strain amplitude of 5. Model fits/predictions are shown for the HAWB, AAB, and AKM models. The data show top/bottom asymmetry, left/right asymmetry, and a vertical shift indicating the presence of thixotropy and viscoelasticity. Figure adapted from Horner et al., Journal of Rheology, 2018,[52] reprinted with permission.

transmittance or backscattering can be measured. When RBCs are more strongly aggregated, the sample will let transmit more light as there will be void regions depleted in RBCs in the sample, yielding a similar aggregation index for these measurements. Light transmittance or backscattering measurements have also been used under dynamic conditions.[70, 71]

While microscopy and light transmission measurements are useful for characterizing the sample as a whole, other methods enable probing the aggregation strength directly between RBCs. These techniques involve bringing two RBCs into contact through various *in vitro* techniques then measuring the disaggregation force. Methods for holding two RBCs in close contact include micropipette aspiration,[72] optical trapping,[73] and atomic force microscopy (AFM).[74] The contributing effects to the aggregation and disaggregation potentials between adjacent RBCs can also be measured through direct techniques. For example, the repulsive surface charge between RBCs has been measured through electrophoretic mobility experiments.[75] Additionally, the deformability of RBCs affects the aggregation of RBCs and contributes to the overall flow behavior at high flow rates. The deformability can be measured under stasis using micropipette aspiration,[76] filtration,[77] or optical tweezers[78] and under dynamic conditions using ektacytometry – a technique based on laser diffraction that can infer both the deformability and orientation of RBCs.[79]

## Constitutive blood modeling

While the non-Newtonian character of blood rheology has been recognized scientifically fairly early, more than 100 years ago[2, 11] almost in parallel with the development of the science of rheology,[80, 81] still, until perhaps the first decade of the second millennium, many of the blood flow dynamics simulations assumed, for convenience, a Newtonian behavior.[82] However, since then, ever more works show that, even when the predominant shear rates are high, when one anticipate non-Newtonian effects to be minimized, significant non-Newtonian effects are observed.[83]

From the four Non-Newtonian characteristics of blood rheology, namely shear thinning, yield stress, thixotropy and viscoelasticity, the first two can be easily captured through the use of a variety of generalized non-Newtonian equations—see Yilmaz and Gundogdu[6] for an exhaustive enumeration of possibilities. Of these, a special mention here is the historic Casson model (Eq. (1))[47], both because this was the workhorse model for describing blood viscosity,[21, 84] but primarily because it is the one for which we have the most extensive knowledge of the parametric dependence of the model parameters on important physiological parameters, such as the hematocrit and the fibrinogen concentration.[38] In its more natural form the Casson model is cast as a linear dependence of the square root of the shear stress with respect to the shear rate,[47] giving rise to the following expression for the equivalent generalized shear viscosity $\eta(\dot{\gamma})$.

$$\eta(\dot{\gamma}) = \left(\left(\frac{\sigma_{y,c}}{|\dot{\gamma}|}\right)^{\frac{1}{2}} + \eta_{\infty}^{\frac{1}{2}}\right)^2. \tag{2}$$

As $\dot{\gamma} \to 0$ the viscosity approaches infinity, and typically approximations are required when used within computational simulations.[85]

As mentioned, an extra appeal of the Casson model is that several expressions are available in correlating its model parameters with physiological parameters, starting with the pioneering work of Merrill.[86, 87] The most complete is the one developed by Apostolidis and Beris[38] that parametrizes the Casson model viscosity and yield stress in terms of the hematocrit ($H$), fibrinogen concentration ($c_f$), and temperature ($T$):

$$\frac{\eta_\infty}{\eta_p} = (1 + 2.07H + 3.72H^2)\exp\left(-7.03\left(1 - \frac{T_0}{T}\right)\right), \tag{3}$$

$$\sigma_{y,c} = \begin{cases} (H - H_c)^2(0.508c_f + 0.452)^2, & H > H_c, \\ 0, & H \le H_c \end{cases} \tag{4}$$

$$H_c = \begin{cases} 0.313c_f^2 - 0.468c_f + 0.176, & c_f > 0.75 \\ 0.0012, & c_f \le 0.75 \end{cases}, \tag{5}$$

where $T_0$ is the reference temperature at which the plasma viscosity is measured and $H_c$ is a critical hematocrit. The hematocrit is represented as a fraction between 0 and 1. Note that some of the constants in this model have been rounded to 3 significant digits. Additionally, the units for yield stress are dyn/cm$^2$, the units for fibrinogen are g/dL, and the units for temperature are K. More recently, an attempt has been made to incorporate the effects of cholesterol and triglyceride levels into this model.[39] based on an early experimental investigation by Moreno and coworkers.[42] An early application of the parametrization of the model parameters on the local hematocrit has been through coupling with a popular stress migration model[88] to represent hematocrit margination in the continuum simulations of blood flow through microvascular bifurcations, which showed a good agreement to microscopic simulations.[89]

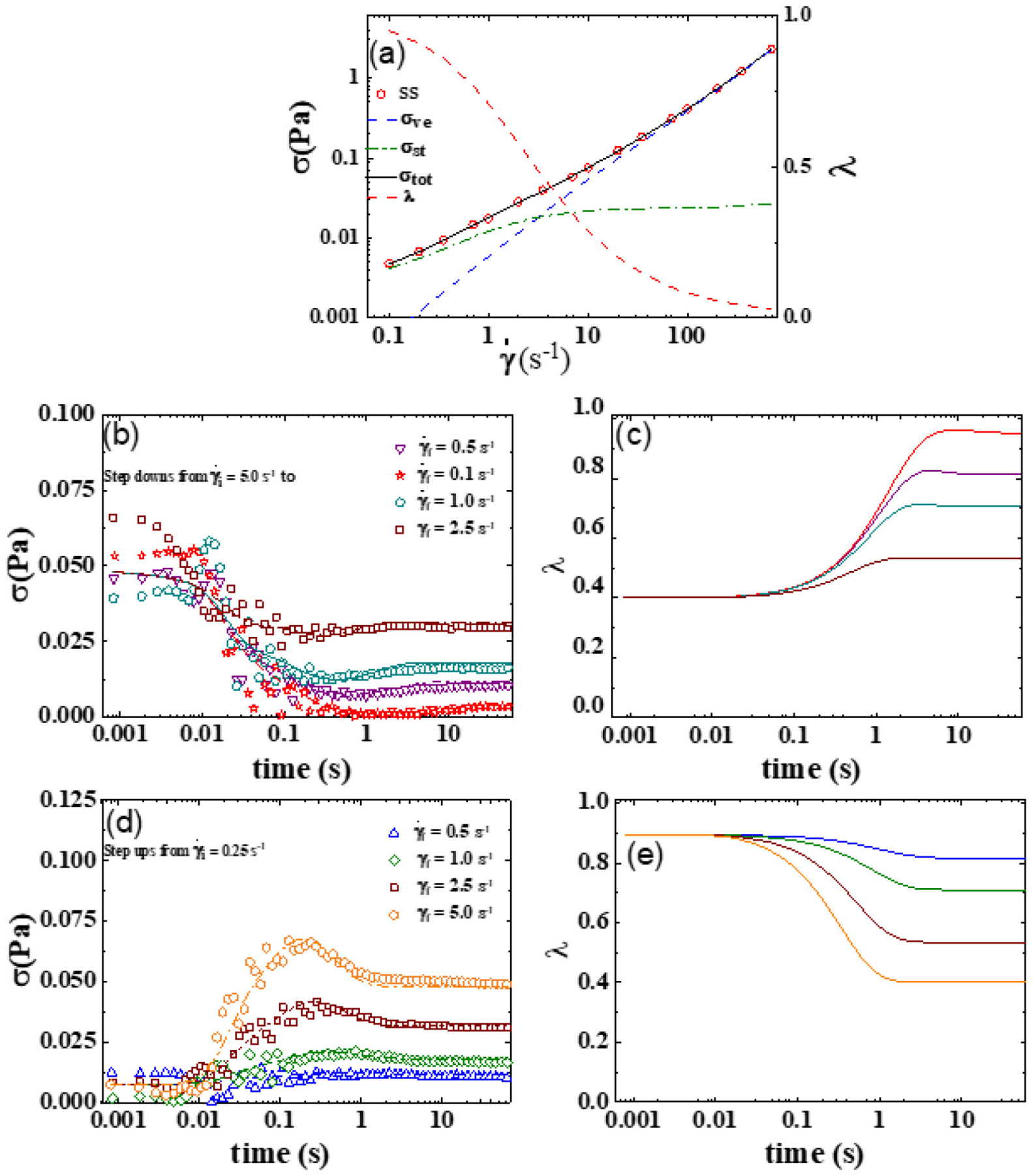

**Fig. 7** Enhanced thixotropic-mHAWB model fit (with lines) compared with total stress data (represented with symbols) for (a) steady state human blood data. The dot dash is the elastic contribution, and the long dash is the viscous contribution. The short dash represents the structure parameter to be read on the right axis. (b) 4 step down transients from initial shear rate of $5.0s^{-1}$ to 0.1, 0.5, 1.0 and $2.5s^{-1}$; (c) corresponding structure parameter curves (colors of structure parameter curves match respective colors from stress evolution curves); (d) 4 step up transients from initial shear rate of $0.25s^{-1}$ to 0.5, 1, 2.5 and $5s^{-1}$; and (e) corresponding structure parameter curves. Reproduced from Armstrong and Tussing, Physics of Fluids, 2020,[108] with the permission of AIP Publishing.

However, a major limitation until the development of the above parametrization has been the scarce availability of data on well characterized blood samples and the absence of a uniform experimental protocol. A major breakthrough in that respect has been the validation of the previously proposed data collection protocol[32] and further elaboration on a systematic steady but also transient shear experimental protocol by Horner et al.[49] and the availability of a set of well documented experimental data on both human and several animal species blood[90]—see also separate section in this review on "Comparative hemodynamics and Hemorheology." This recent work has led to more accurate data that not only advances our knowledge about non-Newtonian blood rheology in *transient* flows, which is not captured of course through inelastic, generalized Newtonian approaches, but also suggests small, but important, corrections to the Casson model predictions and the square root law even regarding the steady state shear viscosity.[13, 52] –see **Fig. 5** and the following section.

At this point it is worth mentioning earlier fundamental work on this subject following a viscoelastic modeling approach, such as, Campo-Deaño et al.[60], Shariatkhah et al.,[91] where the authors used the Giesekus and Phan-Thein and Tanner models, as well as Rajagopal and coworkers who developed thermodynamically based models.[92, 93]

Nonetheless, it may still be necessary to correlate model parameters with blood physiochemistry extending beyond what has been provided to date (i.e., hematocrit, fibrinogen, cholesterol levels) as additional factors may still be at play such as the blood type,[87] oxygenation levels[94] and of course, aging[95] and a variety of pathological conditions.[96] Regarding the latter, the relevant literature is growing at a hefty pace,[97] a clear indication of the emergent importance of blood rheology as a diagnostic tool. The following modeling focuses on physiologically healthy conditions, which will be a necessary baseline for any future use of hemorheology as a diagnostic aid.

## Thixotropic/Elastoviscoplastic models for blood

Generalized Newtonian models are computationally inexpensive to implement but cannot predict accurately transient (thixotropic/hysteretic) changes in the viscosity which are relevant as blood flows naturally under pulsatile conditions. To better account for the transient effects, a viscoelastic model can be used. One example of this is the Anand-Kwack-Masud (AKM) model which is a generalization of the Oldroyd-B model:

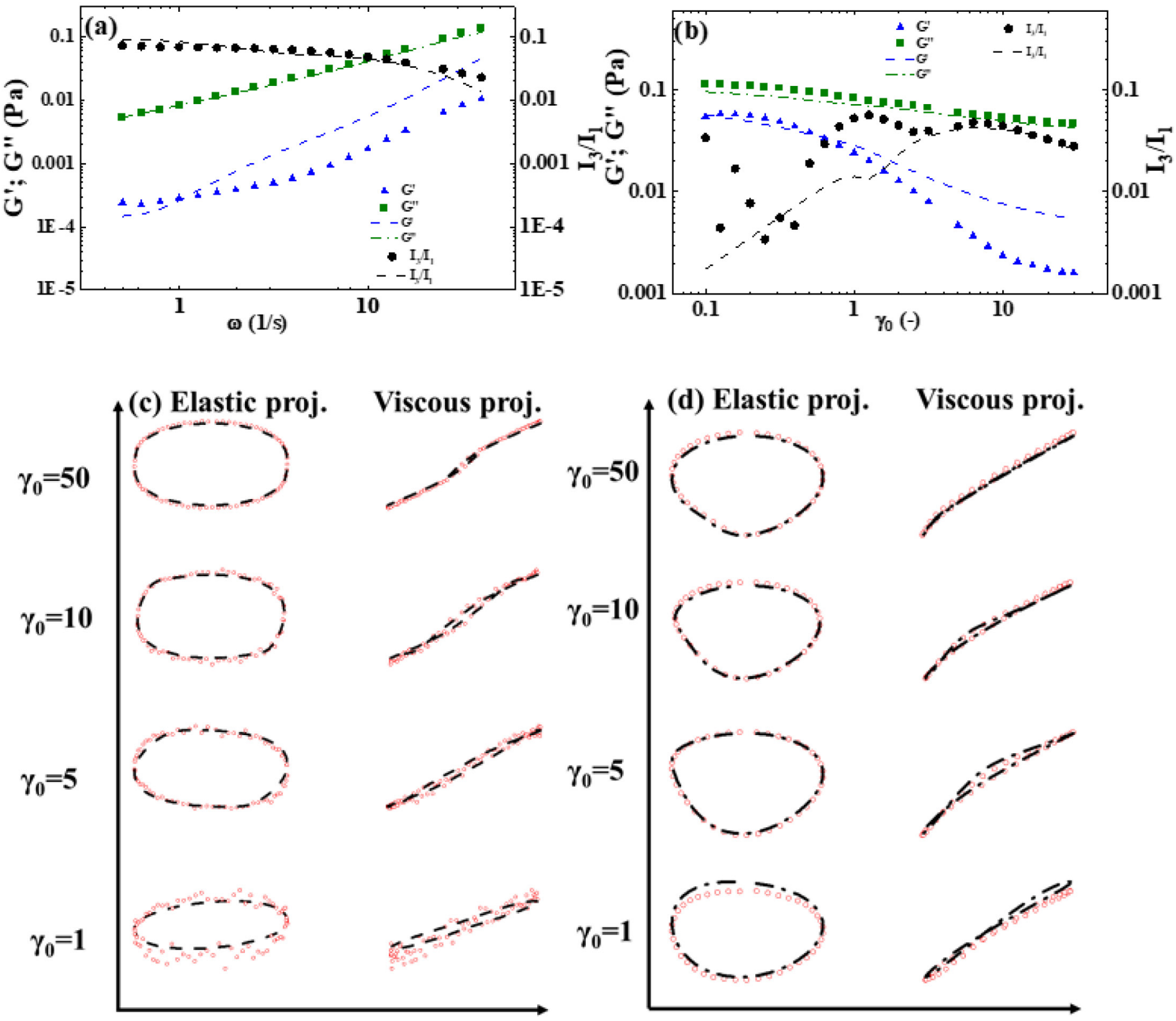

**Fig. 8** Enhanced thixotropic-mHAWB model predictions (with lines) compared with experimental data (represented with symbols) (a) Frequency sweep at =10(-) and (b) Amplitude sweep at =12.566 (rad/s) (c) Elastic and viscous LAOS projections, showing the one cycle of oscillatory stress vs strain and strain rate respectively at $\omega = 1$ rad/s; $\gamma_0 = 1,5,10,50$; (d) Effective elastic and viscous UDLAOS projections, showing one cycle of stress vs strain and strain rate at $\omega = 1$ rad/s; $\gamma_0 = 1,5,10,50$. Reproduced from Armstrong and Tussing, Physics of Fluids, 2020,[108] with the permission of AIP Publishing.

$$\boldsymbol{\sigma} = G\mathbf{B} + 2\eta_\infty \mathbf{D}, \tag{6}$$

$$\overset{\nabla}{\mathbf{B}} = \left(\frac{G}{\eta_0 - \eta_\infty}\right)\left(\frac{\mathrm{tr}(\mathbf{B})}{3}\right)^m \left(\mathbf{B} - \frac{3}{\mathrm{tr}(\mathbf{B}^{-1})}\mathbf{I}\right), \tag{7}$$

where $\boldsymbol{\sigma}$ is the stress tensor, $G$ is an elastic modulus, $\mathbf{B}$ is the left Cauchy-Green stretch tensor, $\mathbf{D}$ is rate of strain tensor, the superimposed inverted triangle ($\nabla$) denotes the upper-convected time derivative, $m$ is a power law index, and $\mathbf{I}$ is the unit tensor.[98] Viscoelastic models similar to this approach are advantageous for modeling blood flow as they can account for elasticity in the sample and can predict non-shear components of the stress tensor. However, they are typically unable to capture thixotropic effects that blood demonstrates associated with the structural evolution.

Various authors have proposed thixotropic models to describe the transient rheology of blood.[99-102] These models typically rely on a kinetic rate equation for a parameter which describes structure within the system. This parameter is then connected to a description describing the contribution to the stress by the rouleaux structures. Notable in this respect, in its pioneering description that unifies thixotropy and viscoelasticity, is a model by Sun and De Kee that uses a generalized Maxwell viscoelastic model with viscoelastic parameters that depend on a structure parameter governed by a separate evolution equation.[103] Much more recently, Stephanou and Georgiou,[102] employed a similar approach to derive variants of those equations that are compatible with the general equation for non-equilibrium reversible-irreversible coupling (GENERIC) formalism[104] and non-equilibrium thermodynamics.[105]

One of the most successful thixotropic models for blood is the Horner-Armstrong-Wagner-Beris (HAWB) model[52] which uses this approach to describe the rouleaux contribution to the shear stress $\sigma_{R,12}$ as:

$$\sigma_{R,12} = \lambda^3 \eta_R \dot{\gamma}_p + \lambda G_R \gamma_e, \tag{8}$$

$$\frac{d\lambda}{dt} = \frac{1}{\tau_\lambda}\left((1-\lambda) + (1-\lambda)\tau_a|\dot{\gamma}_p| - \lambda\left(\tau_b|\dot{\gamma}_p|\right)^2\right), \tag{9}$$

where $\lambda$ represents the level of microstructure (rouleaux formation) from [0,1], where 0 represents absence of structure, i.e., full individual RBCs, and 1 fully formed rouleaux, $\lambda^3$ represents the best (for the HAWB) power-law dependence of the rouleaux to an effective viscosity with $\eta_R$ as the viscosity of fully formed rouleaux, $G_R$ is the yield stress modulus, $\gamma_e$ is the elastic strain, $\tau_\lambda, \tau_a, \tau_b$ are three characteristic times for Brownian aggregation, shear-induced aggregation and shear-induced breakage, respectively, and $\dot{\gamma}_p$ is the plastic strain rate. We note here that in this modeling paradigm the strain and strain rate are decomposed into plastic and elastic components thusly: $\gamma_{tot} = \gamma_e + \gamma_p$, and $\dot{\gamma}_{tot} = \dot{\gamma}_e + \dot{\gamma}_p$. The mathematics of this construct, following the kinematic hardening theory of plasticity,[106] were originally introduced by Dimitriou et al.[107], and modified by Apostolidis et al.[101] and Horner et al.[13, 52] The elastic component is utilized to describe reversible processes, whereas the plastic component is incorporated in modeling irreversible processes with respect to the aggregation and breaking apart of the rouleaux. The final equations for the HAWB model are provided as follows.[52]

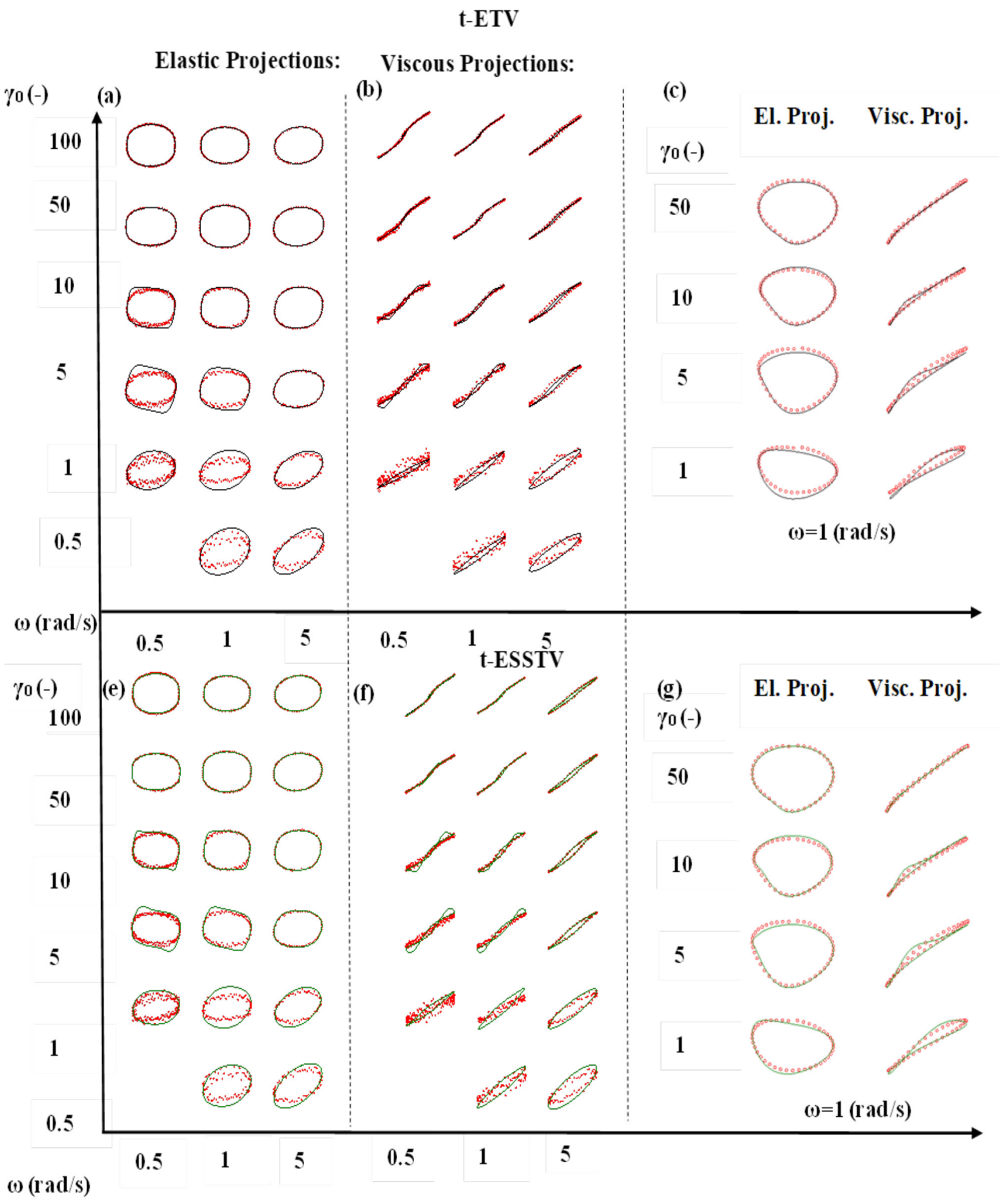

**Fig. 9** Elastic and viscous LAOS (a-b, e-f) and UD-LAOS (c,g) projections for t-ETV (a-c) and t-ESSTV (f-g) models. The elastic projections are shown as oscillatory stress over one cycle with respect to the oscillatory strain, whereas the viscous projections are shown as oscillatory stress plotted against oscillatory strain rate. The elastic projections are shown for frequencies, $\omega$ and strain amplitude $\gamma_0$ values as indicated. See also Armstrong et al., Journal of Rheology, submission no.: JOR21-AR-00184.[118]

The evolution equation for the elastic strain rate is given by:

$$\dot{\gamma}_e = \begin{cases} \dot{\gamma}_p - \frac{\gamma_e}{\gamma_{max}}|\dot{\gamma_p}|, & \frac{d\gamma_{max}}{dt} \geq 0 \\ \dot{\gamma}_p - \frac{\gamma_e}{\gamma_{max}}|\dot{\gamma_p}| + \frac{\gamma_e}{\gamma_{max}}\frac{d\gamma_{max}}{dt}, & \frac{d\gamma_{max}}{dt} < 0 \end{cases}, \tag{10}$$

in which the maximum elastic strain sustained by the rouleaux, $\dot{\gamma}_{max}$, is defined as:

$$\dot{\gamma}_{max} = \gamma_{0,R}\lambda. \tag{11}$$

The specific plastic component of the rate of strain on the rouleaux can be evaluated by

$$\dot{\gamma}_p = \begin{cases} \frac{\dot{\gamma}}{\left(2 - \frac{\gamma_e}{\gamma_{max}}\right)}, & \dot{\gamma} \geq 0 \\ \frac{\dot{\gamma}}{\left(2 + \frac{\gamma_e}{\gamma_{max}}\right)}, & \dot{\gamma} < 0 \end{cases}. \tag{12}$$

To further improve the accuracy of the blood rheology model at higher shear rates, it has been necessary to include in addition to the rouleaux stress contribution mentioned above a separate contribution to the stress tensor $\boldsymbol{\sigma}_C$ arising from the individual red blood cells and the plasma. In the original HAWB model this was introduced as a viscoelastic stress $\boldsymbol{\sigma}_C$ that was defined through an extended White-Metzner model[52] as

$$\boldsymbol{\sigma}_C + \left(\frac{\eta_C(\boldsymbol{\sigma}_C)}{G_C}\right)\overset{\nabla}{\boldsymbol{\sigma}}_C = \eta_C(\boldsymbol{\sigma}_C)\dot{\gamma}, \tag{13}$$

where $\eta_C$ is so defined as a function of the first invariant of the stress tensor $\sigma_{I,C}$, in such a way as to reduce to Cross viscosity[52] fit of the shear thinning behavior due to individual RBCs (as shown in **Fig. 1**(c) for hardened RBCs).[5, 10]

The original HAWB, and its subsequent modifications are the only models with the ability to unify viscoelasticity and thixotropy, which was shown to be critical when modeling transient blood rheology.[52] A comparison of the original HAWB model, the AKM model, and the Apostolidis-Armstrong-Beris (AAB) model[101] is shown in **Fig. 6** for human blood under Unidirectional Large Amplitude Oscillatory Shear (UD-LAOS) conditions.[52]

Despite the success of the HAWB model, it had several limitations, some of which were quickly addressed in subsequent modifications. First, Horner et al.[13] constructed the mHAWB framework with a reduced $\lambda$ evolution equation without the shear-induced aggregation and with a modified shear-induced breakage, where the nondimensional structure parameter, $\lambda$, evolves according to a kinetic rate equation:

$$\frac{d\lambda}{dt} = \frac{1}{\tau_\lambda}\left((1-\lambda) - \lambda\tau_b|\dot{\gamma}|\right), \tag{14}$$

where $\tau_\lambda$ and $\tau_b$ are time constants governing the overall structure evolution and the rate of structure breakup, respectively. This reduces the number of parameters without affecting the accuracy of the fits. Second, it introduced a viscoelastic contribution from the Rouleaux in lieu of a viscous one, that is found from the following equation newly incorporated structural viscoelastic stress ($\sigma_{ve,R}$), is defined according to a Maxwell-like equation with a structure dependent modulus and viscosity:

$$\frac{d}{dt}\sigma_{ve,R,12} = \begin{cases} G_R\lambda^{1.5}\left(\dot{\gamma}_p - \frac{\sigma_{ve,R,12}}{\eta_R\lambda^{1.5}}\right), & \frac{d\lambda}{dt} \geq 0 \\ G_R\lambda^{1.5}\left(\dot{\gamma}_p - \frac{\sigma_{ve,R,12}}{\eta_R\lambda^{1.5}}\right) + 1.5\frac{\sigma_{ve,R,12}}{\lambda}\frac{d\lambda}{dt}, & \frac{d\lambda}{dt} < 0 \end{cases}, \tag{15}$$

where $G_R$ and $\eta_R$ are the fully structured structural modulus and viscosity, respectively.

The evolution of models based on the mHAWB framework all contain three key components: 1. A viscoelastic contribution to total stress from the individual RBCs, and plasma; 2. a viscoelastic contribution to total stress from the rouleaux; and 3. a microstructure (rouleaux) evolution equation that takes into account shear breakage, shear aggregation (only for the HAWB version of the model) and Brownian aggregation.[13, 52, 108-110].

In parallel to this model evolution, significant work has been devoted to investigating the main aspects of macroscopic blood rheology modeling: a) regarding the viscoelastic modeling of the individual cell deformation viscoelasticity and then applied this 'enhanced thixotropy version' to several viscoelastic models of the Oldroyd-8 family and the Giesekus model (Table A, B)[108, 109] b) regarding the contribution to the total stress form the rouleaux, and c) regarding the thixotropic structural parameter evolution and a comparison against a well-known elasto-visco-plastic framework used by Varchanis et al.[111]

Armstrong and Tussing[108] went back to a classic thixotropy evolution framework involving three items: 1. shear breakage term; 2. shear aggregation term; and 3. a Brownian aggregation term; with the moniker, enhanced thixotropy. With a slightly modified structural evolution equation, albeit similar to the original HAWB structural evolution equation:

$$\frac{d\lambda}{dt} = \frac{1}{\tau_\lambda}\left((1-\lambda) + tr_2(1-\lambda)|\dot{\gamma}_p|^m - tr_1\lambda|\dot{\gamma}_p|\right), \tag{16}$$

where $tr_1$ is the overall time constant of shear breakage, and $tr_2$ is the overall time constant of shear aggregation with $m = 1/2$, based on best literature values for human blood[108, 109]. We believe ethixo-mHAWB is one of the best models we have so far to represent the complex rheology of blood. The quality of fits using the ethixo-mHAWB model for steady state, step down and step up in shear rate can be seen in **Fig. 7**, and the quality of predictions of the model for LAOS and UD-LAOS are shown in **Fig. 8** (adapted from Armstrong and Tussing[108]—see also the same reference for more information).

Using the identical equations for elastic and plastic strain and shear rate as shown above, along with the enhanced thixotropic expression for structure parameter, Armstrong and coworkers demonstrated that this framework could be applied to the Oldroyd-8 and Giesekus families of viscoelastic equations. The structural contributions to total stress from the rouleaux are then linearly superimposed with the contributions from each of the Olrdoyd-8 models. This framework allows for stress predictions in the $xx$, $yy$, and $zz$ direction directions as well as $yx$[108, 109] (See Table 1 in Armstrong and Tussing[108], and Table 1 in Armstrong and Pincot[109]).

Additionally, Armstrong and coworkers tested variants of the AAB model adding more terms to describe the physics of the structural evolution equation, as well as a structural contribution to total stress, with a variant that also contained a viscoelastic time constant for this contribution to total stress from the rouleaux[112] (See Table1 in Armstrong et al.[112]). Moreover, Armstrong and coworkers also interrogated other thixotropic model frameworks from Wei et al., and Varchanis et al.[110, 111, 113] These frameworks, following the original work by Oldroyd[114] and Saramito[115] included the elastic pre-yield viscoplastic behavior within a combined tensorial viscoelastic framework. They also rely heavily on original work from Dimitriou et al.[116], and theories of plasticity, and yielding. [106, 110-112] A similar effort was made by Giannokostas et al.[117] However, in both these works[110, 117], the viscoelastic contribution to total stress from the plasma and individual RBCs, a key feature of the mHAWB construct, was absent.

Considering the recent model developments mentioned above, three more very recent model extensions of the mHAWB model have emerged that are worth mentioning. [118] In all three, only the modeling of the rouleaux contribution to the stress has been modified keeping the modeling of the viscoelastic contribution of the RBC deformation and plasma the same, as provided by equation (13). First, following Varchanis et al.[111] the enhanced structural stress thixo-viscoplastic model (ESSTV)[118] was developed where the total rouleaux contribution to the stress is modeled following

$$\frac{1}{G_R\lambda}\frac{D}{Dt}\sigma_{R,12} + \max\left(0, \frac{\left|\sigma_{R,12} - \left(\frac{\sigma_{y,0}}{\gamma_{0,R}}\right)\gamma_e\right|}{\eta_R\lambda^m}\right)\operatorname{sign}(\sigma_{R,12}) = \dot{\gamma}. \tag{17}$$

Second, this equation has been extended into a full tensorial description, t-ESSTV,[118] as

$$\frac{1}{G_R\lambda}\frac{D}{Dt}\boldsymbol{\sigma}_R + \max\left(0, \frac{\left|\boldsymbol{\sigma}_R - \left(\frac{\sigma_{y,0}}{\gamma_{0,R}}\right)\gamma_e\right|}{\eta_R\lambda^m}\right)\mathrm{sign}(\boldsymbol{\sigma}_R) = \dot{\gamma}. \qquad (18)$$

Moreover, the original viscoelastic rouleaux contribution in the mHAWB model, called in this work extended thixo-viscoelastic model (ETV), has also been cast in a tensorial form, t-ETV model.[118] Most interestingly, this was enabled a fully thermodynamically compatible[105] form using an internal conformation tensor $\mathbf{C}$, which evolves as

$$\overset{\nabla}{\mathbf{C}} = -\left(\frac{G_R\lambda^m}{\eta_R} + \frac{m}{2\lambda}\left(\frac{d\lambda}{dt} + \left|\frac{d\lambda}{dt}\right|\right)\right)(\mathbf{C} - \mathbf{I}), \qquad (19)$$

and is connected to the viscoelastic contribution to the stress as

$$\boldsymbol{\sigma}_V = G_R\lambda^m(\mathbf{C} - \mathbf{I}), \qquad (20)$$

maintaining the same elastic contribution.[118] As shown there[118], there is significant improvement moving from ETV to ESSTV (approximately 10% reduction of the fit dimensionless total residuals), and another 10% improvement gained by moving to the full tensorial, t-ESSTV, model framework. To further judge the quality of the improvement in model predictions **Fig. 9** is provided.

Based on the provided information it is apparent that significant progress has been made in modeling blood behavior that has resulted in successfully identifying and modeling what we believe are the three most dominant components in any macroscopic blood stress modeling: the elastic and viscoelastic contributions arising from the rouleaux structures, and the viscoelastic contribution due to the free cells deformation and possibly plasma (e.g. see recent works attesting on plasma viscoelasticity[37]).

A remaining challenge for the future remains to complete the full tensorial description of the constitutive modeling, including the evolution equation for the structural parameter, and to do that in a way that it is a) full compatible to non-equilibrium thermodynamics and b) in a form that works for all types of flows and not just for shear flow. Additionally, we feel that it will be beneficial to bring in a more direct connection to the evolving microstructure—some effort towards that goal is described in a subsequent section of this review. Moreover, it is worth mentioning here, that the comprehensive models developed for human blood may also be easy to adapt to be used with general animal blood—an effort along that direction and some early experimental findings are also discussed in a subsequent section of this review. Finally, one should not underestimate the complexity of blood involving not only the few factors that have entered as parameters in some of the current models (most notably the hematocrit) but also many more, as many studies have anecdotally shown to be important: various pathological factors, food, stress (cytokines), age, sex, blood type, gases, etc. It therefore becomes obvious that any model parameters need to be carefully adjusted to govern anyone particular case—this has already been clearly shown in our own data sample dependencies 8-10 and is expected to be more enhanced when broader samples are used beyond healthy individual donors. The need for experiments and personalized measurements cannot be therefore underemphasized.

## Microscopic/mesoscopic models of blood

In reality, the inherent complexity of blood, can only be captured well through microscopic modeling. Blood, not only involves a concentrated suspension of deformable, elastic, and aggregating RBCs, but also a host of other ingredients from platelets and WBCs to various proteins.[119] Moreover, all of these interact in a complicated way. This complicated structure is the reason for blood's complex rheology, not only characterized by a non-zero yield stress and shear thinning but also pronounced history effects, i.e., thixotropy, as fully discussed above. Several efforts have therefore been made to reconstruct the non-Newtonian characteristic of blood from micromechanical models based on first principles.[17, 120-122] However, due to the as yet poorly understood biological interactions between its ingredients, a full *a priori* construction of blood rheology remains elusive. Adjustable parameters to describe RBC behavior are still necessary in microscopic as well as mesoscopic level just as at the macroscopic level discussed above.

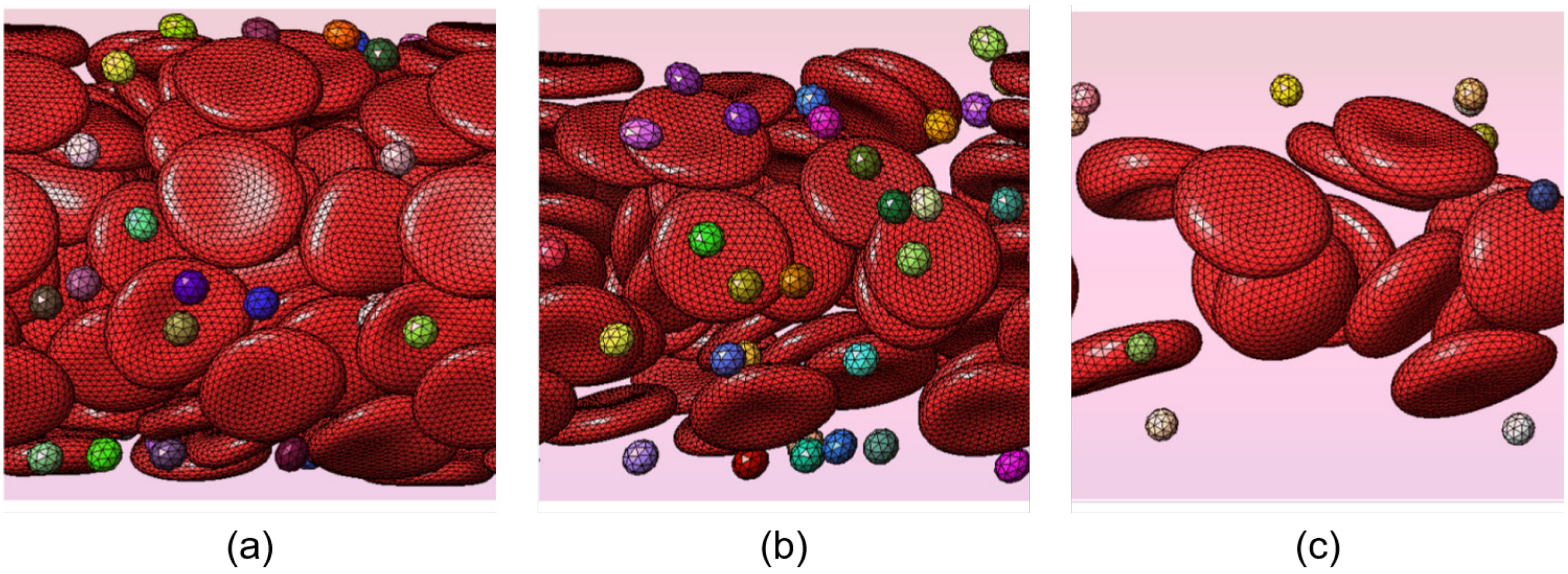

**Fig. 10** Snapshots of numerical simulations of blood flow with RBCs and platelets in a cylindrical channel for three different values of $H$: (a) 40 %, (b) 20 %, (c) 10 %. Blood plasma is modeled with the DPD method. It occupies the remaining volume of the channel (not shown). Figure taken from Bessonov et al., Mathematical Modelling of Natural Phenomena, 2014.[128] Reprinted with permission.

Nevertheless, microscopic and/or mesoscopic simulations have been particularly useful in resolving important details in the flow. For example, it is significant that detailed DPD simulations can produce predictions for steady shear viscosity that from first principles,[17] based on whether RBCs are allowed to aggregate or not, can reproduce almost exactly the experimental observations of Chien et al.[5] shown in **Fig. 1**(c). Other example, such investigation on RBC aggregation[123-125] and especially wall effects,[126-129] namely in explaining the presence of a depletion layer next to the wall surface,[130] are also worth noting, see also **Fig. 10**.[128] As discussed above, these depletion layers are important in flows through small vessels and lead to the well-known Fåhræus[33] and Fåhræus-Lindqvist[34] effects, which have been further analyzed and quantified in the literature.[131] Microscopic simulations have been used to estimate the latter effect[126] as well as to study margination[15] and aggregation[132] in capillary flows. Together with the full analysis of the endothelial glycocalyx, and the wall elasticity, such phenomena are essential when one tries to model blood flow in the microvasculature[133] –see also a recent review.[134]

More recently, simulations became much more sophisticated, for example, taking into account in order to model the aggregation process the role of fibrinogen and even including effects of pathological conditions like type 2 diabetes mellitus[135] and hyperviscosity based on cell interactions, cell stiffness and hematocrit.[136] Especially important is the use of microscopic/mesoscopic models in the evaluation of blood hemostasis and thrombosis – see recent reviews on the subject.[137-139] Of interest is the work on this subject by Yazdani et al.[140] with explicit analysis of the role of the platelets and their shear-dependent adhesive dynamics coupled to the biology of thrombin formation. Of particular interest in association to this (and similar works[139]) is the need for a connection of the micro-meso and macrostructure, which characterizes multiscale approaches, an item that is dealt more specifically separately in the next section.

Typically, all microscopic methods start with an explicit representation of the larger independent structures in blood flow, RBCs, and perhaps other cellular components, such as platelets[141] and WBCs. Those are described at various levels of detail, from the most coarse-grained (spheres)[123] to the more elaborate ones through a detailed elastic network representation of the shape (biconcave disc for RBCs) allowing also for flow deformations.[124, 142] The latter contains an intermediate discrete representation of the RBCs as an ensemble of particles with prescribed forces between them being in the middle[128, 132] – see also the recent reviews.[121, 143, 144] In all cases, the forces between the resolved particles and between the particles and the flow have next to be specified (which therefore also necessitates a representation of the flow). This can be achieved through a Lagrangian approach to develop the resolved particles trajectories by solving the corresponding Newton's equations of motion. Several methods have been developed for that: (1) Using boundary integral methods that can be developed in the limit when inertial effects can be neglected.[142] (2) Using the immersed boundary approach, the fluid equations are solved in the entire domain using with a Lattice Boltzmann (LB) approach,[145, 146] (albeit any computational fluid dynamic (CFD) method, such as finite differences can be employed),[147, 148] with special constraints applied where the resolved particles are to properly enforce particle-flow interactions, following the original work of Peskin.[149] (3) Using a fluid particle-based formulation also for the Navier-Stokes fluid equations[17, 121, 122, 132] – several fluid particle-based methods have been followed over the years, with more emphasis on dissipative particle dynamics (DPD)[17, 121, 122, 150] and its variants, such as the fluid particle model[132] and smoothed DPD.[151]

There are several benefits that can so far be extracted from microscopic blood flow simulations. First, there is theoretical confirmation on the inhomogeneities generated in the flow due to the finite particle sizes and their elastic properties, interparticle interaction forces (leading also to aggregation), and wall exclusion and hydrodynamic forces effects.[142, 145-148, 151] Second, they can generate important detailed information necessary in building and validating coarser scale models. Third, they can eventually make the connection to the underlying biology.[140] The insight gained through these detailed calculations is invaluable and they are definitely needed when special effects are to be modeled (e.g., thrombosis) and/or flow in small capillaries, especially when nanoparticles are involved and inhomogeneous effects dominate.[16] On the other hand, there is the realization about the difficulty of the task originating on the problem complexity and the intimate dependence of the microscopic simulation results on information, such as the role played by different proteins, and more specifically on biological/biochemical reactions, that is either as yet fully unknown or cannot be easily personalized to particular situations.[152] Moreover, with increased complexity come increased computational requirements for the simulations, considerably limiting their applicability realtime to solving everyday problems. Thus, albeit we believe that microscopic simulations play an important role in understanding the blood flow problem, we need to acknowledge that they provide just one piece of the whole puzzle that ultimately requires to involve modeling at all levels: from molecular (and genetic) to microscopic, mesoscopic, and macroscopic. Some of the current efforts in building up true multiscale approaches are reviewed in the next section.

## Multiscale models

Multiscale models attempt to bridge the gap between the nanoscale, mesoscale and macroscale dynamics while retaining relevant information at smaller length scales. Systems such as blood, multicomponent, involving a concentrated suspension of deformable RBC dominated by strong interactions between themselves and several of the multitude of the other present components, have such a degree of complexity that makes a multiscale approach particularly advantageous. Before selecting the approach for multiscale modeling, it is important to consider the flow regime, geometry as well available computational resources. One can either resolve the flow-induced inhomogeneities using a model that couples flow descriptions at multiple length scales, or one can choose to coarse grain some aspects of the flow physics. The latter approach allows one to use fewer computational resources for a trade-off in terms of accuracy and microscopic consistency.

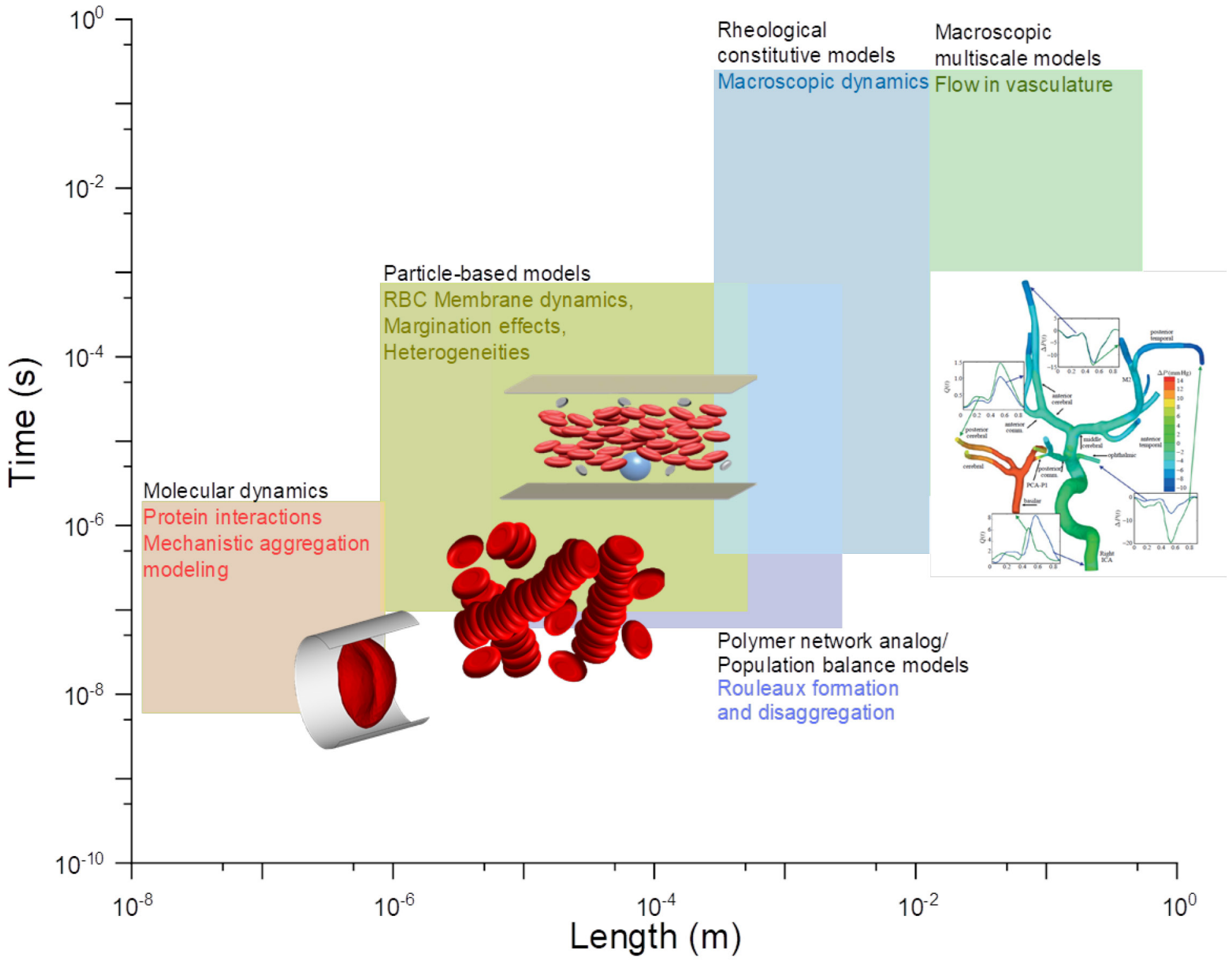

**Fig. 11** Schematic map of length and time scales involved in blood flow modeling. The highlighted areas indicate the range of time scales that can be handled by the methodology with illustrations indicating the modeled physics. Figure adapted from Fedosov et al., Biophysical Journal, 2010;[120] Horner et al., Journal of Rheology, 2018;[52] Kumar and Graham, Soft Matter, 2012;[15] and Grinberg et al., Philosophical Transactions of the Royal Society A, 2009.[155] Reprinted with permission.

Significant progress has been made in the past in increasing the efficiency of multiscale numerical simulations involving the modeling of interactions at different length scales by exploiting graphic processing units (GPU) hardware and parallel message passing (MPI) software[153, 154] advances of parallel computing. This has paved the path forward to directly employ multi-physics solvers that address different length and timescales involved in a problem and exchange information during their run-time. In a physical sense, the problem is typically divided into its continuum and discrete aspects, for example, the vasculature is decomposed, depending on the size of blood vessels, into regions where blood can be modeled as continuum (macroscopic), and where it exhibits and strong inhomogeneities (mesoscopic) or capillary flow (microscopic),[15, 129, 155-157] see also **Fig. 11**. Perdikaris et al.[158] provide an overview of key advancements in the simulation of blood through such multi-physics solution schemes in human arterial trees. Because the modeling of membrane elasticity and vesicle dynamics become computationally intractable for large number of interacting cells, hierarchical simulations typically embed smaller microlevel domains into the larger continuum domain while exchanging information through overlapping patches. Because there is a likelihood of error from the randomness and non-reproducibility of discrete particle-based methods such as DPD, replicates can be performed to get better statistics. Current multi-physics software can be linked to perform simulations under physiological flow conditions in patient-specific vascular geometries. This hybrid approach has been used in a few studies[155-157] to incorporate flow effects at the microscale and modeling physiological phenomena such as brain aneurysm, clot and thrombus formation.[140] The interaction of blood and its constituents with the elastic walls of blood vessels also becomes important while simulating the vasculature, therefore, the flow modeling can incorporate fluid-structure interaction.[129, 159]

At the mesoscopic level one resolves the flow and wall-induced inhomogeneities but employs continuum models rather than microscopic ones. There have been several efforts, particularly, the multimode viscoelastic models developed initially for blood by Owens and coworkers.[100, 160, 161] These models connect the constitutive models and microscopic models by relating the evolving structure to directly observable parameters. The original model achieved this by tracking the aggregation and disaggregation rate of RBCs and linked this to the bulk behavior through a polymer network theory analog.[100, 160] The model was later refined to account for inhomogeneities driven by stress-induced migration.[161, 162] Tsimouri et al.[163] improved this approach by reformulating the governing equations to be consistent with nonequilibrium thermodynamics,[104, 105, 164] based off work originally developed for rodlike micellar solutions[164]. This reformulation avoids numerous arbitrary parameters, enhancing the predictive capabilities of the model.

A similar work was proposed by Jariwala et al.[165] using a population balance model (PBM) that accounts for the effect of rouleaux structure formation by assuming that the red blood cells form low dimensional self-similar fractal structures. This model removes much of the empiricism in determining aggregation-disaggregation behavior by incorporating kinetics scaling grounded in colloidal physics. The model assumes spherical symmetry in rouleaux and works with the volume fraction of red blood cells as a metric for structure formation, thus providing a connection to the blood hematocrit. It adopts a Smoluchowski kernel[166] for aggregation kinetics that provide the size dependent aggregation behavior through Brownian motion and shear. Further, by employing a physically-based PBM within a colloidal framework, the model provides a rational pathway for including more physical phenomena affecting RBC interactions without empiricism.

Additional mention needs to be made to the use of the theory of interacting continua for mixtures which treats blood as a two fluid mixture containing plasma, treated as a viscous fluid, and RBCs, treated as a concentrated suspension with hematocrit and shear rate dependent viscosity, with special modeling for the drag and lift forces.[167] This model has shown to have reasonable agreement with experiments conducted in a sudden expansion microchannel flow.[167]

## Comparative hemodynamics and hemorheology

Because of the frequent use of animals in clinical drug testing, the topic of comparative differences in hemodynamics and hemorheology is of great interest. Across species, blood typically consists of the same makeup with differences in the constituent sizes and concentrations. However, in comparison to order of magnitude changes in the animal sizes, the changes in blood constituent sizes are relatively small and do not show any trend with body mass. As an example, horse RBCs have a mean corpuscular volume (MCV) of approximately 37-59 fL with a hematocrit of 32-53 %; whereas, mouse RBCs have an MCV of approximately 47.4-58.9 fL with a hematocrit of 42.7-56.3 %, despite a roughly 4 order of magnitude difference in body mass between the two species.[168] RBCs from most mammals have the same biconcave disc shape, shown for canines in **Fig. 12**(a),[168] that human RBCs have, with minor exceptions like camel RBCs which have an ellipsoidal shape, shown in **Fig. 12**(b).[168] Mammals are unique to vertebrates in that their RBCs do not contain any organelles; whereas, all other vertebrates have nucleated RBCs, shown for chickens in **Fig. 12**(c).[168] The biconcave disc shape is also a requirement for RBC aggregation into rouleaux, which affect the low shear viscosity of blood.[169] However, due to the fairly consistent hematocrit, RBC size, and plasma viscosity, the whole blood viscosity at high shear rates is relatively constant across species.[170]

In a very recent study, Horner et al.[90] present detailed measurements of the rheology of blood from seven different animal species obtained under identical conditions, involving both steady state and various shear transients, to those used in previous human blood investigations.[13, 52] In addition, they also provided fits against the previously developed mHAWB model.[13] This allowed for a detailed comparison. In general, the fact that the same model, originally developed from human blood studies, was able to fit also all the animal data showed that qualitatively a similar rheological behavior was also observed from all animal species. Therefore, animal blood rheology is also characterized by a shear thinning, viscoelastic and thixotropic behavior. However, quantitatively, the results were different as various model parameters showed significant, species-dependent, differences, occasionally being even outside the range of values encountered in human blood investigations.[90] The differences were more pronounced in the description of a viscoplastic behavior characterized by a non-zero yield stress. More specifically, such a behavior was found to characterize only animal blood known to involve RBC aggregation.[90] In the following we will focus on further discussing the implications of this observation.

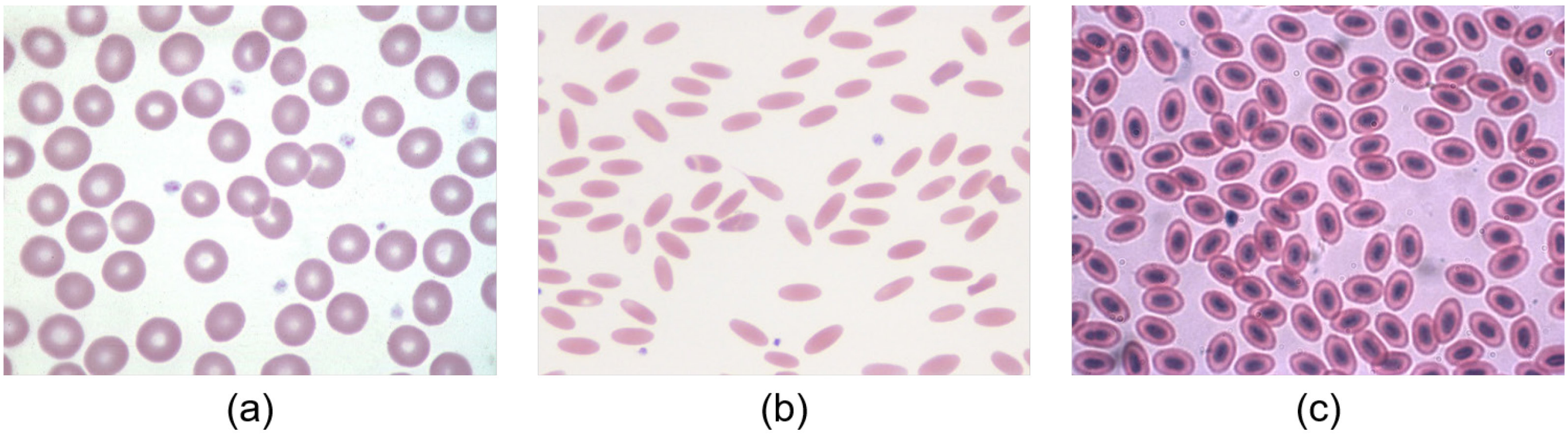

**Fig. 12** RBC images for (a) canine blood, (b) camelid blood, and (c) chicken blood. Figures taken from Cornell University College of Veterinary Medicine eClinPath Hematology Atlas.[168] Reprinted with permission.

Many hemodynamic characteristics across species typically scale with body mass according to a power law relation. This observation gave rise to the field of allometry which has been applied widely to both plants and animals.[171] West and coworkers proposed a landmark model originally for mammals then later expanded to all organisms which attempts to describe this scaling by maximizing the metabolic capacity for a given organism.[172, 173] Shortly following West and coworkers' proposition, the model was heavily disputed, primarily centered around the assumption and interpretation of the space filling fractal nature of blood vessels.[174-176] Nevertheless, the model agrees well with experimental allometric exponents for various physiologically relevant parameters including aorta radius, Womersley number, capillary density, and total resistance.

One major assumption that West and coworkers made when deriving their allometric scaling model was that blood was a homogeneous, Newtonian fluid. However, as presented in the previous sections, blood flow throughout the circulatory system is neither homogeneous nor Newtonian. Similar to human blood, blood across species is also non-Newtonian, although interestingly, the extent of shear thinning that blood exhibits can vary significantly with species, as shown in **Fig. 13**.[90] This is because the extent of RBC aggregation also varies significantly across species. At rest and low shear rates, RBCs from some species like horse, human, and pig will form rouleaux, whereas these structures are absent for other species like cow, goat, and sheep. This observation raises two outstanding questions: What physically causes rouleaux to be present in some species and absent in others? And why evolutionarily did this phenomenon develop in some species?

To address the first question of what causes rouleaux to form in certain blood samples, we can look to the physical properties of the blood. For this analysis, horse and sheep are used as representative species with strong RBC aggregation and weak to no RBC aggregation, respectively. The two main determinants of rouleaux formation in human blood are the hematocrit and the fibrinogen concentration, with higher levels of both leading to more significant rouleaux formation. These factors were explicitly studied by Windberger and coworkers for various species, and although horse displayed far more RBC aggregation than sheep, the horse blood sample used contained a very similar hematocrit value (horse: 35 %, sheep: 33 %) and a lower fibrinogen value (horse: 155 mg/dL, sheep: 282 mg/dL) than the sheep blood sample.[177] Furthermore, Baskurt and coworkers confirmed that the difference in fibrinogen concentration cannot explain the increased aggregation for horse RBCs by resuspending horse RBCs in dextran solutions and comparing the aggregation index to human and rat RBC data. They found that even when resuspended, horse RBCs demonstrated increased aggregation tendency.[178]

Another major factor that affects RBC aggregation is the RBC surface charge. As RBCs are repulsive, a higher surface charge would lead to less pronounced aggregation. The RBC surface charge for various species was measured by Eylar and coworkers, and despite a strong aggregation tendency, horse RBCs showed the highest surface charge density for the species considered, exceeding sheep RBCs (horse: 4230 esu/cm$^2$, sheep: 3610 esu/cm$^2$).[179] It is widely accepted that the deformability of the RBCs affect the viscosity of the blood at high shear rates, but it has also been suggested that the deformability of the RBCs could play a role in the aggregation tendencies. This was also investigated for various species by Plasenzotti and coworkers through ektacytometry. It was determined that horse RBCs showed a higher elongation index than sheep RBCs at high stresses, but at low stresses, the elongation index for horse RBCs was lower than sheep RBCs.[180] This observation is again far from conclusive for determining the difference in extent of aggregation between the RBCs from the two different species. Other authors have proposed that changes across species in the glycocalyx geometry,[20] RBC membrane composition,[181] or fibrinogen structure[182] could all result in the observed variations in RBC aggregation. However, these hypotheses are limited because of the lack of available data across species.

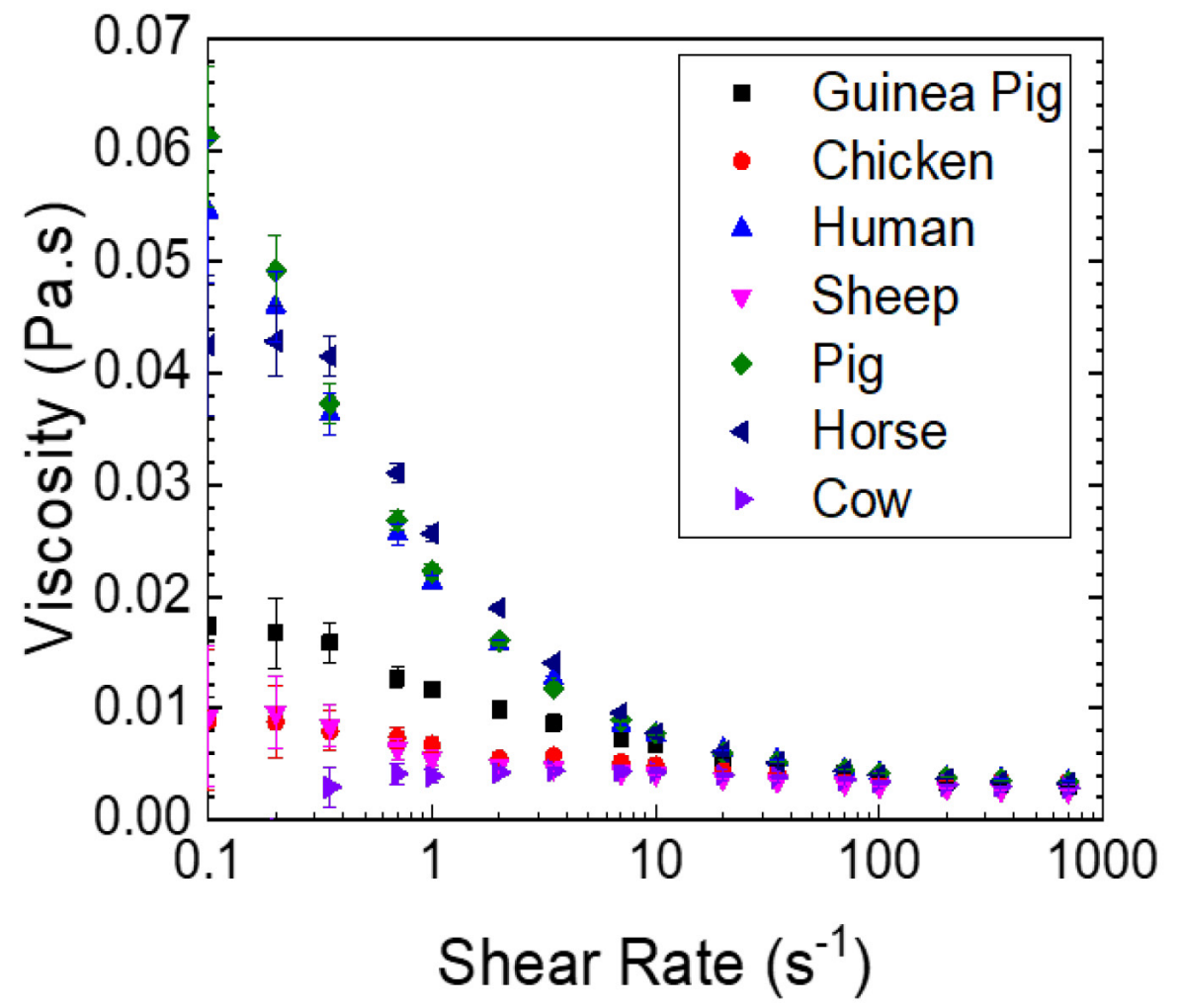

**Fig. 13** Steady shear viscosity for blood from various species. Note that blood from some species shows pronounced shear thinning, while blood from other species appears almost Newtonian. Figure adapted from Horner et al., Soft Matter, 2021.[90]

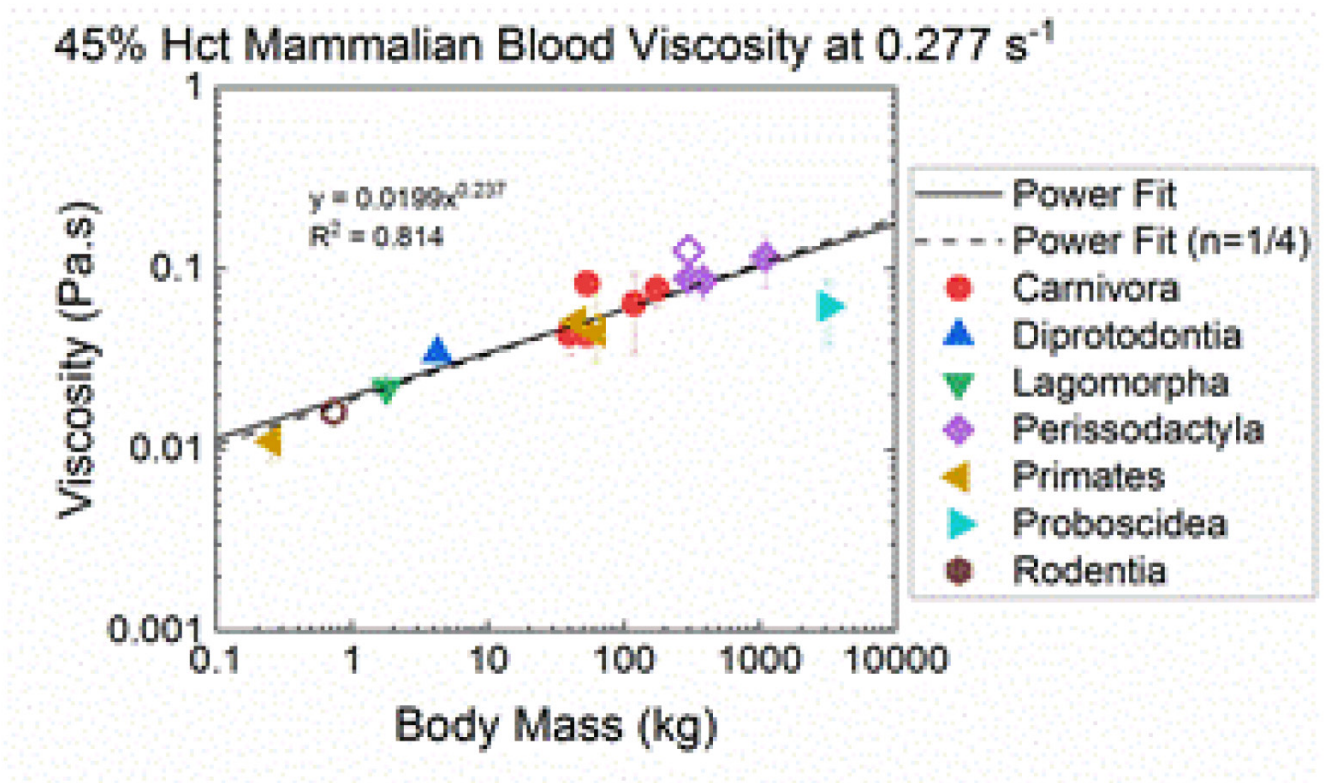

**Fig. 14** Whole blood viscosity at 0.277 s$^{-1}$ and 45% hematocrit as a function of body mass for various mammalian species. Viscosity data are from Johnn *et al.*, 1992;[170] Bodey and Rampling, 1998;[183] and Horner *et al.*, 2021[90] and body mass values are from AnAge: The Animal Ageing and Longevity Database. Figure adapted from Horner et al., Soft Matter, 2021.[90]

The other question regarding the differences in RBC aggregation across species pertains to why evolutionarily these changes arose. One way to address this question is to look to allometry to see if there is a correlation between body mass and aggregation tendency. This assessment was performed by Horner and coworkers, and a correlation between low shear blood viscosity and body mass was identified after omission of artiodactyla, as shown in **Fig. 14**[90, 170, 183]. The viscosity at low shear rates is an indirect indicator of the extent of RBC aggregation. This experimentally observed allometric scaling relation was hypothesized to facilitate a maintenance of wall shear stress across species. Other authors have suggested that the differences in RBC aggregation across species could correlate with the aerobic capacity of the species, with more athletic species demonstrating higher RBC aggregation. A correlation was observed for the species considered, namely antelope, cow, dog, horse, and sheep.[184] However, as pointed out by other researchers, this correlation fails when more species, like rodents which are very active but have minimal RBC aggregation, are included.[185] Other evolutionary factors that could dictate how the material properties of blood have evolved could include diet, native environment, or hemostasis requirements. As of this date, however, there is no clear evidence that any variation in these factors correlate well with the aggregation ability of RBCs across species.

The significance of rouleaux formation across species for Poiseuille flow, which is more representative of *in vivo* conditions, has been studied sparingly. Fåhræus conducted experiments of capillary flow of cattle, human, and horse blood.[30] He observed that cattle blood demonstrated a more homogeneous flow profile than that of human or horse blood. Human RBCs under flow, as expected, gave rise to a depletion layer near the walls of the capillary, while horse RBCs formed dense clusters in the tube with large depletion layers throughout. From this, it was postulated that the increased RBC aggregation could be advantageous by increasing the cell free marginal layer and correspondingly decreasing the apparent viscosity. However, at the same time, increased RBC aggregation could lead to obstructions in smaller vessels downstream.[30] Although this observation could potentially explain some of the differences in RBC aggregation across species, a more comprehensive study is still required to confirm this. In general, there are many open questions in comparative hemodynamics and hemorheology, which are difficult to answer due to the large number of factors that change across species and the dearth of experimental measurements on blood from different species. Moreover, the protocol used for the existing studies often varies from one author to another making it difficult to compare results. Thus, more comprehensive research into this subject is required.

## Hemorheology, disease and diagnostics

A better, more quantitative understanding linking the flow properties of blood to composition and variations in red-blood cell morphology and properties has significant diagnostic potential.[96] Diseases, such as hyperviscosity,[136] hypertension,[186] sickle cell anemia, [187, 188] and diabetes,[135, 189] have been linked to changes in the viscosity of blood. Blood rheological measurements at the point of care can provide a rapid diagnostics tool as well as a method for preventative medicine indicating early signs of such diseases. Furthermore, hemorheology is essential for a more quantitative understanding of blood flow in vivo,[190] which in turn can provide insight into undesired conditions such as thrombus formation[137-140] and provide guidance on treatment strategies.[191, 192] The combination with microrheological investigations of individual blood cells has been proposed as a method for personalized medicine.[193] Furthermore, thickening of blood rheology has been correlated with aging effects, where regular exercise has been shown to be beneficial to improving overall health in the elderly by improving hemorheology.[95]

## Summary and outlook

Despite having studied hemodynamics for over 2000 years, we are only just starting to understand the complex rheological nature of blood. In this review, several recent notable contributions have been highlighted in both experiments and modeling and placed into the proper context in association to earlier seminal works. We discussed the complex material properties of blood – namely pseudoplasticity, viscoelasticity, and thixotropy – and how these arise from the constituent interactions. Closely related to these phenomena, are the complex flow phenomena that also arise for in vivo and in vitro conditions, such as the Fåhræus and Fåhræus-Lindqvist effects. It was also discussed how hemorheology varies across species, leading to a new allometric scaling with body mass, and how these interspecies variations can be advantageous for a specific organism. On the modeling side, various recent advances were discussed which were separated into constitutive and microscopic/mesoscopic & multiscale modeling approaches. For constitutive modeling, the evolution of models ranges from traditional generalized Newtonian models to modern multimode models that can incorporate viscoelasticity and thixotropy.

Accurate predictions of steady and transient blood viscosity are demonstrated by models that account for the thixotropy and viscoelasticity arising from rouleaux formation as well as the viscoelasticity developed by stretching the RBCs at higher shear rates. Modern modeling methods, incorporating hierarchical modeling, also show promise, and their significance in understanding biological interactions and inhomogeneous effects in blood flow has been highlighted.

Because of these recent advances, we are beginning to uncover the clinical benefits of understanding hemodynamics and hemorheology. However, there are several areas that are not well understood. Particularly, a more complete understanding of the relation between biology, physiology and hemorheology can improve current clinical diagnostic tests, like the erythrocyte sedimentation rate (ESR) test, to not only detect infections but be able to differentiate diseases and even identify patients at risk for developing cardiovascular complications. Moreover, by better understanding how the material properties of blood change when removed from the body and how this relates to the numerous biological reactions occurring, we can improve storage guidelines for extracted blood samples. Another future direction for this field is understanding how blood responds in extreme situations, such as low gravity, high altitude, or under intense exercise. Measurements under these conditions could lead to improved health screening and establishment of better guidelines, which could prevent cardiovascular complications. Importantly, the recent discovery that the SARS-CoV2 virus infection in humans is actually a cardiovascular infection raises critically important concerns about the hemorheology of the disease state for treatment as well as potential rapid diagnosis.[194]
Additionally, it could help to determine the optimal hematocrit for peak human performance. This discovery could be instrumental in the development of artificial blood as it could be designed it to have the optimal flow properties.

From the modeling perspective, there is also vast opportunity in the field of hemodynamics. One of the major problems still remaining when developing efficient numerical simulations of blood flow of the arterial vascular tree is the specification of appropriate boundary conditions to express surface-flow interactions and their computational stability in connection with a complex blood rheology and wall deformation characteristics. By improving both the physical accuracy of the models used and the accuracy of the numerical schemes implemented for their efficient solution, while decreasing the computational workload, perhaps through innovative parallel implementations of new multiscale and person-specific approaches, we could begin to work towards developing a full, accurate model for the circulatory system. This could have widespread benefits ranging from a better understanding of drug delivery to improved surgical and transfusion techniques. Another area of interest moving forward, is to evaluate whether hemodynamic and hemorheology modeling principles can be applied to other biofluids. Several other clinically relevant biofluids, such as synovial fluid, saliva, mucus, and vitreous humor, exhibit interesting material properties. Research into these fluids tends to be even sparser than blood. By better understanding the rheology of multiple biofluids and in multiple species in general, we can learn a significant amount about how these fluids evolved to meet their specific needs.

## Conflicts of interest

There are no conflicts to declare.

## Acknowledgements

The authors acknowledge the financial support from the National Science Foundation through CBET 1510837.

## Nomenclature

**Latin symbols**

| | |
|---|---|
| $a$ | size (radius) of red blood cell [m] |
| $c_f$ | Fibrinogen [g/dL] |
| $\mathbf{B}$ | left Cauchy Green stretch tensor [-] |
| $\mathbf{C}$ | conformation tensor |
| $\mathbf{D}$ | rate of strain tensor [$\mathrm{s}^{-1}$] |
| $G$ | shear elastic modulus [Pa] |
| $G_C$ | shear elastic modulus for extended White-Metzner contribution [Pa] |
| $G_m$ | shear modulus of RBC membrane [Pa m] |
| $G_R$ | Shear elastic modulus associated with rouleaux [Pa] |
| $G'$ | shear storage modulus [Pa] |
| $G''$ | shear loss modulus [Pa] |
| $H$ | Hematocrit |
| $H_c$ | critical hematocrit |
| $\mathbf{I}$ | identity tensor |
| $k_B$ | Boltzmann constant [$\mathrm{J\ K}^{-1}$] |
| $m$ | shear aggregation exponent |
| $r$ | radius [m] |
| $t$ | time [s] |
| $tr_1$ | overall shear breakage time constant [s] |
| $tr_2$ | overall shear aggregation time constant [s] |
| $T$ | temperature [K] |
| $T_0$ | reference temperature [K] |

**Greek Symbols**

| | |
|---|---|
| $\dot{\gamma}$ | shear rate [$\mathrm{s}^{-1}$] |
| $\gamma_0$ | shear strain amplitude [-] |
| $\gamma_e, \dot{\gamma}_e$ | elastic shear strain [-], elastic shear rate [$\mathrm{s}^{-1}$] |
| $\gamma_p, \dot{\gamma}_p$ | plastic shear strain [-], plastic shear rate [$\mathrm{s}^{-1}$] |
| $\gamma_{max}$ | maximum elastic strain [-] |
| $\nabla$ | upper convected time derivative |
| $\eta$ | Viscosity [Pa s] |
| $\eta_C$ | plasma/ individual RBC contribution to viscosity [Pa s] |
| $\eta_{st}$ | structural contribution to viscosity [Pa s] |
| $\eta_0$ | zero-shear viscosity [Pa s] |
| $\eta_\infty$ | infinite-shear viscosity [Pa s] |
| $\eta_p$ | plasma viscosity [Pa s] |
| $\eta_R$ | rouleaux viscosity [Pa s] |
| $\rho$ | fluid density $[\mathrm{kg\ m^{-3}}]$ |
| $\sigma$ | Stress [Pa s] |
| $\boldsymbol{\sigma}$ | stress tensor [Pa s] |
| $\boldsymbol{\sigma_R}$ | rouleaux stress contribution [Pa s] |
| $\boldsymbol{\sigma_C}$ | Individual RBC stress contribution [Pa s] |
| $\boldsymbol{\sigma_V}$ | viscoelastic rouleaux contribution [Pa s] |
| $\sigma_{ve,R}$ | viscoelastic rouleaux contribution to shear stress [Pa s] |
| $\sigma_y$ | yield stress [Pa s] |
| $\sigma_{y,c}$ | Casson model yield-stress [Pa s] |
| $\tau$ | time constants [s] |
| $\tau_a$ | time constant for shear induced aggregation relative to that due to Brownian motion [s] |
| $\tau_b$ | time constant for shear induced breakage relative to that due to Brownian motion [s] |
| $\tau_\lambda$ | time constant for Brownian aggregation [s] |
| $\lambda$ | structure parameter [-] |
| $\omega$ | angular/rotational velocity $[\mathrm{rad\ s^{-1}}]$ |

## Notes and References